\documentclass[a4paper,aps,superscriptaddress,floatfix,nofootinbib,twocolumn]{revtex4-1}

\usepackage{bbold}
\usepackage{bbm}
\usepackage[pdftex]{graphicx}
\usepackage{latexsym,amsmath,verbatim,amssymb,txfonts}
\usepackage{color}
\usepackage{rotating}
\usepackage{verbatim}
\usepackage{multirow}
\usepackage[english]{babel}
\usepackage{comment}

\newcommand{\change}[1]{#1}

\begin{document}

\title{Error-Correcting Decoders for Communities in Networks}

\author{Krishna C. Bathina}
\affiliation{Center for Complex Networks and Systems Research, School
  of Informatics, Computing, and Engineering, Indiana University, Bloomington,
  Indiana 47408, USA}

\author{Filippo Radicchi}
\affiliation{Center for Complex Networks and Systems Research, School
  of Informatics, Computing, and Engineering, Indiana University, Bloomington,
  Indiana 47408, USA}
\email{filiradi@indiana.edu}

\begin{abstract} 
As recent work demonstrated, the task of identifying communities in networks 
can be considered analogous to the classical problem of
decoding messages transmitted along a noisy channel. 
We leverage this analogy to develop a community detection 
method directly inspired by a standard
and widely-used decoding technique. We further simplify the 
algorithm to reduce the time complexity from quadratic to linear. 
We test the performance of the original and reduced versions of the algorithm on 
artificial benchmarks with pre-imposed community structure, and on 
real networks with annotated community structure.
Results of our systematic analysis indicate that the proposed 
techniques are able to provide satisfactory results. 
\end{abstract}

\maketitle

\section{Introduction}

Real networks often exhibit organization in communities, 
intuitively defined as groups of nodes with a higher density of edges within
rather than between groups~\cite{girvan2002community, fortunato2010community}. 
Most of the research on this topic has 
focused on the development of algorithms for 
community identification. Proposed approaches vary widely, 
including hierarchical clustering algorithms~\cite{friedman2001elements}, modularity-based
methods~\cite{newman2004finding, newman2004fast, clauset2004finding,
	guimera2007module, duch2005community,
	newman2006finding,newman2006modularity}, 
random walk based algorithms~\cite{
	zhou2003distance, rosvall2008maps}, and statistical inference
methods~\cite{newman2007mixture,hastings2006community,
	decelle2011inference,karrer2011stochastic,peixoto2014hierarchical,
	peixoto2013parsimonious, peixoto2017bayesian},
to mention a few of them. 
Whereas algorithms differ much in spirit, they all share two intrinsic
limitations.  First, as described by the No Free Lunch
Theorem~\cite{peel2017ground}, there is no community detection
algorithm that works 
best for all networks and community structures; 
an algorithm good for one class of networks may be equally bad for another class of networks. 
A second type of limitation arises from self-consistency tests, where community
detection methods are applied to instances of the stochastic block
model to uncover the community structure  pre-imposed in the model.
Algorithms can recover a non-vanishing portion of the true community
structure of the graph  only if the amount of fuzziness in the network
is below  the detectability threshold~\cite{decelle2011inference,nadakuditi2012graph,krzakala2013spectral,radicchi2013detectability,
radicchi2014paradox, abbe2015community,abbe2017community}. Also, exact
detection of the true cluster structure is subjected to a threshold
phenomenon
~\cite{abbe2016exact,abbe2017community,mossel2013proof}. 
This phenomenon can be understood through the lens of coding theory 
by interpreting the problem of defining and identifying communities in networks 
as a classical communication task over a noisy channel, analogous to 
the one originally considered by Shannon~\cite{shannon2001mathematical}.
The value of the exact recovery threshold can be estimated
in the limit of infinitely large graphs~\cite{abbe2016exact,abbe2015community,abbe2017community,mossel2013proof}.
A bound on the value of the threshold for finite-size graphs can  be
obtained as an application of the Shannon's noisy-channel 
coding theorem~\cite{radicchi2017decoding}.

In this paper, we exploit the analogy between coding theory and
community structure in networks, and develop a novel class of algorithms for
community detection based on a state-of-the-art decoding technique~\cite{gallager1962low,mackay1996near}.
The idea has been already considered in Ref.~\cite{radicchi2017decoding} for the simplest case of network
bipartitions. Here, we expand the method to find multiple communities
by iterating the bipartition method in a way similar to what already
considered in
Refs.~\cite{newman2013community,kernighan1970efficient,fiduccia1982linear}. 
As the decoding method considered in Ref.~\cite{radicchi2017decoding}
has computational complexity that scales quadratically with the
\change{number of nodes in the network}, we further propose an approximation of the algorithm that makes
the method complexity scale linearly \change{with the number of
  edges, thus making it linearly dependent with system size in sparse networks}. 
We perform systematic tests of the both algorithm versions on synthetic and
real-world graphs. Performances appear satisfactory in all cases.

\section{Methods}

\subsection{Community detection as a communication process}

\begin{figure}[!htb]
	\centering
	\includegraphics[width=.5\textwidth]{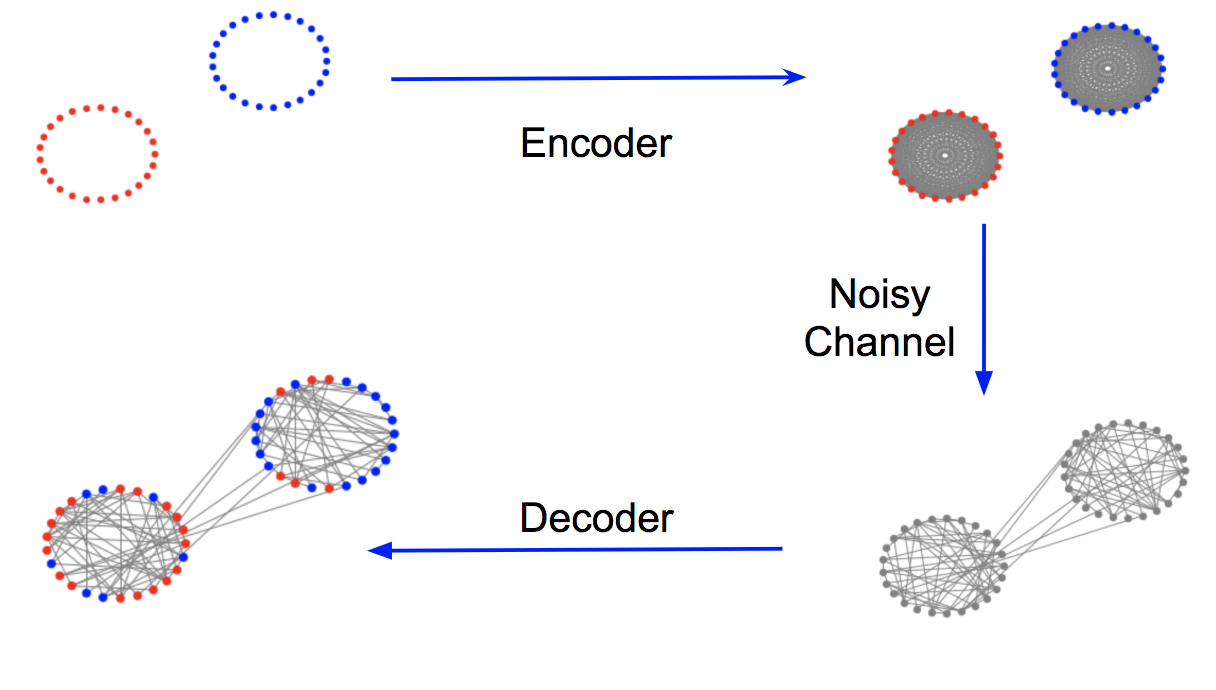}
	\caption{Community detection as a decoding task of a message
          transmitted along a noisy channel.  A message made up of community assignments is formed
		into a network structure through an encoder. The codeword
		is then transmitted trough a noisy channel. The
                channel noise delete any information regarding the
                assignment of nodes to  communities, and further
                deteriorates the network structure by deleting/adding edges. The
                observed network is received at the end of the noisy
                channel, and its structure is used to decode the original message.}
	\label{fig:communication}
\end{figure}

For sake of clarity, we repeat the same description already provided
in Refs.~\cite{abbe2016exact,abbe2015community,abbe2017community,mossel2013proof,
radicchi2017decoding} of how the definition and detection of
communities in a network can be framed as a communication process (see
Figure~\ref{fig:communication}).

We assume that there are $N$ nodes in the network and that each node
$i$ has associated a single information bit $\sigma_i =0, 1$. The
value of the bit identifies the group of node $i$.
 The message is encoded by adding $N(N-1)/2$
parity bits $\theta$, each for every pair of nodes. The parity bit
$\theta_{i,j} =0$ if $\sigma_i = \sigma_j$, or $\theta_{i,j} =1$, otherwise.
The parity bits are essentially added to the original message
according to the rule 
\begin{eqnarray}
\sigma_i + \sigma_j + \theta_{i,j} = 0  \; ,
\label{eq:parity}
\end{eqnarray}
where the sum is performed in modulo-$2$ arithmetic. The set of
$N(N-1)/2$ equations defines the code used in the communication
process. In the jargon of coding theory, Eqs.~(\ref{eq:parity})
  defines a low-density parity-check (LDPC) code. These type of codes
  are
often used in practical communication tasks, given their
effectiveness~\cite{gallager1962low,mackay1996near,
  mackay2003information}. 
In graphical terms, the encoded message can be seen as
a network composed of two disconnected cliques, where each
identifies a community of nodes.

Once encoded, the message is transmitted trough a communication
channel. There, noise alters the bit values. Information bits $\sigma$ are
deleted so that there is no longer information about node
memberships; some parity bits $\theta$ are flipped 
giving rise to the observed network.
The goal of the decoder is to use information from the observed network
together with a hypothesis on the noise characterizing the channel to
infer the original message about group memberships.

\subsection{Stochastic block model as a noisy channel}

As already done in
Refs.~\cite{abbe2016exact,abbe2015community,abbe2017community,mossel2013proof,
radicchi2017decoding}, we make a strong hypothesis on
the noisy channel. We assume that the observed network is given by a
stochastic block model, where pairs of nodes within the same group are 
connected with probability
$p_{in}$, and pairs of nodes belonging to different groups are
connected with probability $p_{out}$. This corresponds to assuming
that the noisy channel is given by an asymmetric binary channel, and
that the parity bits $\theta$ are flipped with probabilities defined
in Table~\ref{table: channel}. 
Further, it allows us to use Bayes'
theorem to derive the conditional probability
$P(\theta_{i,j}|A_{i,j})$ for the parity check bit $\theta_{i,j}$
depending on whether nodes $i$ and $j$ are connected in the observed
network, i.e., $A_{i,j} =1$ or $A_{i,j} =0$. \change{Please note that,
since there is no prior knowledge of
the true parity bits values, we assume $P(\theta_{i,j} = 1) = 1/2$
\cite{radicchi2017decoding}.  This represents a strong assumption in
the model, and the resulting algorithm 
is biased towards the detection of homogenous communities.}

\begin{table}
	\centering
	\caption{The conditional probabilities between for the
          variables $A_{i,j}$ and
          $\theta_{i,j}$. The last column was calculated using Bayes'
          rule \change{ with an assumption on the prior of
            $P(\theta_{i,j} = 1) = 1/2$}.} 
	\label{table: channel}
	\begin{tabular}{cccc}
		\hline
		$A_{i,j}$ & $\theta_{i,j}$ & $P(A_{i,j}|\theta_{i,j})$ & $P(\theta_{i,j}|A_{i,j})$ \\ \hline \\ [-0.5em]
		1         & 0              & $p_{in}$              &   $\frac{p_{in}}{p_{in} + p_{out}}$      \\ \\ [-0.5em]
		1         & 1              & $p_{out}$            &   $\frac{p_{out}}{p_{in} + p_{out}}$ \\ \\ [-0.5em]
		0         & 0              & 1 - $p_{in}$         &   $\frac{1-p_{in}}{2-(p_{in} + p_{out})}$  \\ \\ [-0.5em]
		0         & 1              & 1 - $p_{out}$       &   $\frac{1-p_{out}}{2-(p_{in} + p_{out})}$  \\ \hline   
	\end{tabular}
\end{table}

\subsection{Gallager community decoder}

To find the community structure of an observed network, we take
advantage of a widely-used decoding technique for 
LDPC codes. The technique consists in iteratively solving the system of
parity-check equations that defines the code, given the knowledge of
the noisy channel~\cite{gallager1962low,mackay1996near}.
The application of the method to community detection was 
considered in Ref.~\cite{radicchi2017decoding}. 
Specifically, the technique is used to
solve Eqs.~(\ref{eq:parity}) using properties of the channel from
Table~\ref{table: channel}.
The $t$-th iteration of the algorithm is based on
\begin{eqnarray}
	\label{original}
	\begin{split}
		\zeta_{i \rightarrow j}^{t} &= 
		\begin{cases} 
			\ell_i & t = 0 \\
			\ell_i +  \sum_{s \neq i, j} F[\tanh\frac{\ell_{i,s}}{2},\zeta_{s \rightarrow i}^{t-1} ] & t >  0 
		\end{cases} 
	\end{split}
\end{eqnarray}
for all ordered pairs of nodes $i \to j$. The function $F$ is defined as
\begin{eqnarray}
\label{eq:F}
F[a,x] = \log \frac{1 + a\tanh{\frac{x}{2}}}{1 - a\tanh{\frac{x}{2}}}
\; ,
\end{eqnarray}
where $\tanh(\cdot)$ is the hyperbolic
tangent function. In the algorithm, the quantity $\ell_i$ is the log-likehood ratio (LLR)
$\ell_i = \log P(\sigma_i=0) - \log P(\sigma_i=1)$ associated with
node $i$, that is the natural
logarithm of the ratio between the probabilities that the parity bit $\sigma_i$ equals
zero or one.   $\ell_{i,j} = \log
P(\theta_{i,j} = 0 | A_{i,j}) - \log P(\theta_{i,j} = 1 |
A_{i,j})$  is instead the LLR associated with the parity bit
$\theta_{i,j}$ given the hypothesis on  the noisy channel and 
the evidence from the observed network.
The variable  $\zeta_{i \rightarrow j}^{t} $ is still a LLR. It is defined for all
pairs of nodes $i$ and $j$, irrespective of whether they are connected
or not. $\zeta_{i \rightarrow j}^{t} $ may be interpreted as a message that node $i$ sends to node
$j$ regarding the value that the information bit $\sigma_i$ should
assume based on the knowledge of the code, the noisy channel, and the
evidence collected by observing the network. Please note that two
distinct messages  are exchanged for every
pair of nodes $i$ and $j$, depending on the direction of the message, 
either $i \to j$ or $j \to i$. At every iteration $t$, convergence of
the algorithm is tested by first calculating the best estimates of the
LLRs as
\begin{eqnarray}
	\label{parity-quadratic}
	\begin{split}
		\hat{\ell}_i^t &= \ell_i +  \sum_{s \neq  i} F[\tanh\frac{\ell_{i,s}}{2},\zeta_{s \rightarrow i}^{t-1} ] \\
		\hat{\ell}_{i,j}^t &= \ell_{i,j} + F[\tanh\frac{\zeta_{i \rightarrow j}^{t-1}}{2},\zeta_{j \rightarrow i}^{t-1} ]
	\end{split} \; .
\end{eqnarray}
Then, one evaluates the best estimates of the information 
bits, according to $\hat{\sigma}_i = 0$ if
$\hat{\ell}_i^t > 0$, and $\hat{\sigma}_i = 1$, otherwise. A similar
rule is used for the best estimate of the parity bit
$\hat{\theta}_{i,j}$. Finally, the best estimates 
of the bits are plugged in the system of
Eqs.~(\ref{eq:parity}). If the equations are all satisfied, the
algorithm has converged. 
Otherwise, one continues iterating for a maximum
number of iterations $T$. In our calculations, we set $T=100$.

We remark three important facts.
First, possible solutions of the algorithm are
classifications of nodes in either one or two groups. In the first
case, the algorithm indicates absence of block structure in the network.
Second, knowledge of the noisy channel and evidence of the observed network is
used in the definition of the initial LLRs $\ell_{i,j}$. For the
choice of the initial values  of the LLRs for individual nodes $\ell_i$ there is not a
specific rule. If the community structure is strong enough, initial
conditions for the iterative algorithm are not very important.
However, in regimes where community structure is less neat, they may
determine the basis of attraction for the iterated map.
In this paper, we will consider two different choices for the initial values of
the nodes' LLRs. Finally,  we stress that 
the algorithm is the {\it ad-literam} adaptation of the Gallager
decoding algorithm to the detection of two communities. As such, the algorithm
iterates over all possible pairs of nodes, irrespective of whether
they are connected or not. Each iteration of the algorithm 
requires a number of operations that scales with the network size $N$ 
as $\mathcal{O}(N^2)$, thus making the algorithm
applicable only to small/medium sized  networks. 

\subsection{Reducing the computational complexity of the community decoder}
We leverage
network sparsity to reduce the computational complexity of the
algorithm
without significantly deteriorating algorithm performance. The way we decrease
the complexity is rather intuitive. 
In the original implementation, a node sends a message to all other
nodes, even if there is not an edge connecting them. 
In the reduced algorithm, we instead assume that (i) messages are
delivered only
along existing edges, (ii) the message passed from a node to any 
unconnected node is the same regardless of the actual pair of nodes considered. 
This reduces the total number of messages to twice the number of edges
in the network, and thus the complexity from
$\mathcal{O}(N^2)$ to $\mathcal{O}(N \langle k \rangle^2)$, where
$\langle k \rangle$ is the average degree of the network. Our proposed reduction makes the
algorithm linearly dependent on the \change{number of edges in the
  network, which corresponds to a linear dependence with the system size
  if the network is sparse}. 

Specifically, the equations that define the algorithm are as follows.
For connected pairs of nodes $i$ and $j$, we define the initial message
$\zeta_{i \rightarrow j}^{t =0} = \ell_i $, and 
\begin{eqnarray}
\begin{array}{ll}
\zeta_{i \rightarrow j}^{t} = & \ell_i  +  (N - k_i - 1) \; F\left[\tanh{
	\frac{\ell_{non}}{2}},\mathcal{Z} ^ {t - 1}\right] +
\\
& + \sum_{s \in \mathcal{N}_i
	\backslash j} 
\; F\left[\tanh{\frac{\ell_{con}}{2}}, \zeta_{s\rightarrow i}^{t-1}\right] 
\end{array} \; 
\label{edges}
\end{eqnarray}
for iteration $t \geq 1$. In the equation above, $\ell_{non}$ stands for the LLR of
non-connected node, and $\ell_{con}$ is the LLR for
connected nodes. These quantities are defined as

\begin{eqnarray} 
\label{LLR-reduced}
\begin{split}
\ell_{non} &= \log P(\theta_{i,j} = 0| A_{i,j} =0) - \log
P(\theta_{i,j} = 1| A_{i,j} =0) \\
\ell_{con} &= \log P(\theta_{i,j} = 0| A_{i,j} =1) - \log
P(\theta_{i,j} = 1| A_{i,j} =1)
\end{split} \; .
\end{eqnarray}

Further, in Eq.~(\ref{edges}), $k_i$ is the degree of node $i$, and 
$\mathcal{N}_i$ indicates the set of neighbors of node $i$.
Non-existing edges deliver the single message $\mathcal{Z}$. This corresponds
to  the average value of all 
messages among non-connected pairs of nodes in the original 
version of the algorithm. The equations that define
the iterations for $\mathcal{Z}$ are
\begin{eqnarray}
\mathcal{Z}^{t = 0} = \frac{\sum_{i = 1}^N (N - k_i - 1)\ell_i}{N(N-1)
	- 2M} \
\label{non-edges0}
\end{eqnarray}
and
\begin{eqnarray}
\begin{array}{ll}
\mathcal{Z} ^ {t}  = &
\mathcal{Z}^{t = 0} +  \; F\left[\tanh{
                       \frac{\ell_{non}}{2}},\mathcal{Z} ^ {t -
                       1}\right] + \\
& + \frac{ \sum_{i=1}^N \sum_{j \in \mathcal{N}_i}  \;
  F\left[\tanh{\frac{\ell_{con}}{2}}, \zeta_{i\rightarrow j}^{t-1}\right]}{N(N-1) - 2M}
\label{non-edges}
\end{array}
\end{eqnarray}
for iteration $t \geq 1$. 
We used $2M = \sum_i k_i$,  i.e., the sum of the degrees of all the nodes in the network. 

Convergence of the equations above is tested using
the same procedure described in the original algorithm. In particular, the best estimates of the LLRs are computed using

\begin{eqnarray}
	\label{parity-linear}
	\begin{split}
		\hat{\ell}_i &= \ell_i  + \sum_{s \in \mathcal{N}_i}
		\; F\left[\tanh{\frac{\ell_{con}}{2}}, \zeta_{s\rightarrow
			i}^{t-1}\right] + (N - k_i) \; F\left[ \tanh{\frac{\ell_{non}}{2}},\mathcal{Z} ^ {t - 1}\right] \\
		\hat{\ell}_{i,j} &= \log \frac{p_{in}}{p_{out}} + \; F\left[ \tanh
		\frac{\zeta_{i\rightarrow j}^{t-1}}{2}, \zeta_{j\rightarrow
			i}^{t-1}\right] \; .
	\end{split}	
\end{eqnarray}
These values are used to find the best estimates of the bits $\sigma$s
and $\theta$s and, in turn, are plugged into the parity-check Eqs.~(\ref{eq:parity}).
To keep the computational complexity linear, only
parity-check equations corresponding to existing edges are actually tested.
The maximum number of iterations $T$ that we considered before
stopping the algorithm for lack of convergence is $T=1,000$.

\subsection{Initial conditions}

As we mentioned above, the initial value $\ell_i$ of the LLR for every
node $i$ requires initialization. The initialization is potentially
a very important decision for the performance of the algorithm as it
determines the basin of attraction of the iterative system of equations. 
In this paper, we consider two different strategies for the
determination of the starting conditions:

\begin{description}
	\item[Regular] A random node $i$  is chosen such that 
	$\ell_i = 1$ and $\ell_{j} = 0$,  $\forall j \neq i$.
	\item[Random] For every node $i = 1,
	\ldots, N$, $\ell_i$ is a random variable extracted from the
        uniform distribution with support $[-1, 1]$. 
\end{description}

\subsection{Multiple communities}

\begin{figure}[!htb]
	\centering
	\includegraphics[width=.5\textwidth]{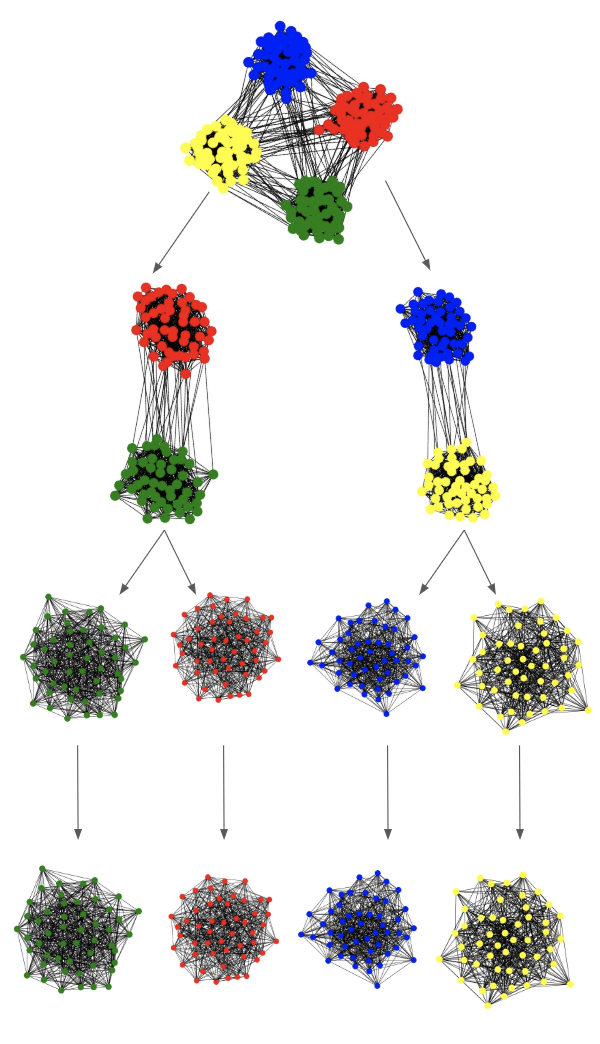}
	\caption{
Schematic representation of the iterative procedure used by the
algorithm to detect multiple communities. In this example, the top graph is a sample network with 4
		communities. In the first iteration, the algorithm splits
		the network perfectly into two equal 
		communities. In the next iteration, each network is split 
		perfectly again. The algorithm terminates because the 
		next iterations do not lead to the discovery of other sub-communities. }
	\label{splits}
\end{figure}

Up to now, we have described how to find a bipartition in a network
according to our procedure. We remark that the output of the algorithm
may also indicate no division of the network. Our goal, however,
is to detect an arbitrary number of communities in our graph. To this
end, we adopt a simple iterative procedure (see Figure~\ref{splits}).
The procedure is identical to the one already adopted in
Refs. ~\cite{newman2013community,kernighan1970efficient,fiduccia1982linear},
and it may be summarized as follows. At the beginning, we define a
list of subgraphs $L$ to be analyzed, and a list of detected
communities $C$. The list $L$ contains only one
element, the entire graph  $G$, while $C$ is empty. 
We then apply the following steps:
\begin{itemize}
	
	\item[1] Take a graph $g$ from the list $L$. Remove the graph from the
	list.
	
	\item[2] Apply the bipartition algorithm to the graph $g$.

          \begin{itemize}
          \item[a]
          If the
	algorithm finds a split of $g$ in two sets of nodes, namely
        $g_1$ and $g_2$, reconstruct each set as a
        graph using only nodes within the set, and only edges
between pairs of nodes within the set. 
        Place $g_1$ and $g_2$ into
        the list $L$. 

        \item[b]
        If the algorithm finds only a set of nodes, so
        that no actual split was detected, $g$ is considered as
a community and placed in the list $C$.
	
\end{itemize}

	\item[3] Go back to point 2 until $L$ is empty. The list of
          detected communities is given by $C$.
	
\end{itemize}

\subsection{Learning the parameters of the noisy channel}

So far, we tacitly assumed to know the values of the probabilities
$p_{in}$ and $p_{out}$. The assumption has been used in the
bipartition algorithm of Ref.~\cite{radicchi2017decoding} when applied
to instances of the stochastic block model with two communities.  In
practical situations, however, prior
knowledge of the probabilities $p_{in}$ and $p_{out}$ is not
available. These parameters should instead be learned in a
self-consistent way by the algorithm relying only on information
from the observed network. Here, we simultaneously propose and validate a 
simple learning strategy.
To this end, we generate instances of the so-called Girvan-Newman
(GN) benchmark graph~\cite{girvan2002community}, a variant of the stochastic model with $N=128$ and
$Q=4$ communities. Different from the original version of the GN model
we allow nodes to have average degree $\langle k \rangle \neq 16$.
The average connectivity of
the model is set by fixing the sum of the true parameter values
$\hat{p}_{in}$ and $\hat{p}_{out}$,  while the strength of the community
structure is instead determined by their difference. We consider four
different combinations $(\hat{p}_{in}, \hat{p}_{out})$ for the true
values of the model parameters to generate four instances of the
model. 
To each of the four instances,  we apply the original algorithm 
with the regular starting conditions to
the network using the parameters values $\tilde{p}_{in}$ and $\tilde{p}_{out}$.              
We measure the
performance of the algorithm to recover the pre-imposed community
structure of the graph, using  
normalized mutual information (NMI) 

\begin{eqnarray} 
	\label{NMI}
	NMI = \frac{I(True, Predicted)}{\sqrt{H(True) H(Predicted)}}
  \; .
\end{eqnarray}

NMI is defined as the mutual information $I$ between the predicted and
true clusters normalized by the square root of the product of the
individual entropies $H$~\cite{strehl2002cluster,
  danon2005comparing}.

\begin{figure}[!htb]
	\centering
	\includegraphics[width=.5\textwidth]{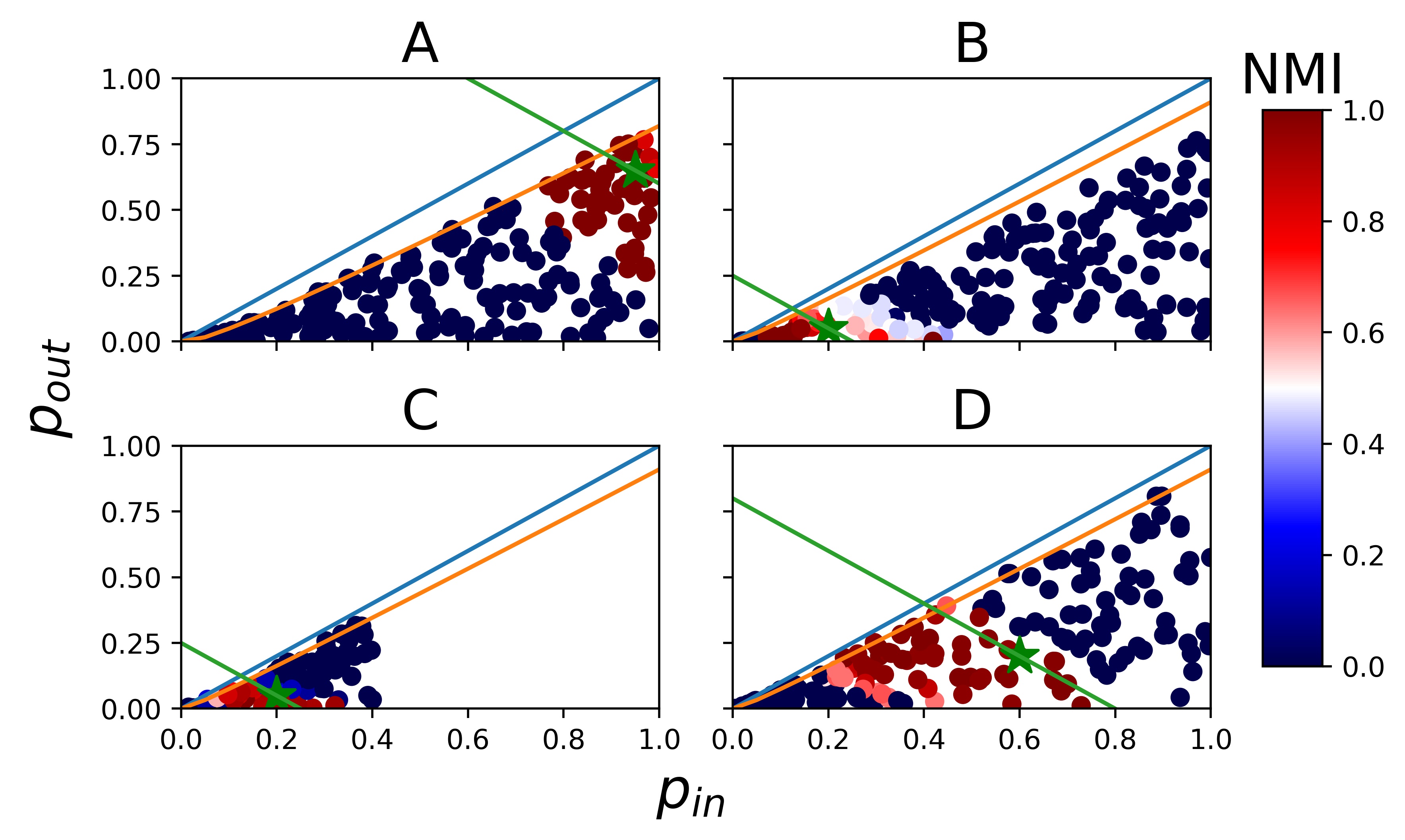}
	\caption{Learning the parameters of the noisy channel. 
Each subgraph report results obtained on an artificial network constructed according to a synthetic model
similar to  the Girvan-Newman benchmark, 
where $N=128$ are divided into $Q=4$ communities
of equal size. Nodes within the same group are connected with
probability $\hat{p}_{in}$, while pairs of nodes belonging to
different groups are connected with probability $\hat{p}_{out}$. We
consider four different combinations $(\hat{p}_{in}, \hat{p}_{out})$
to generate four different instances of the model. Ground-truth values
of $\hat{p}_{in}$ and $\hat{p}_{out}$ are denoted by the green star symbol
in the various panels. We apply the method for community detection
introduced in this paper to the graph using the parameters 
values $\tilde{p}_{in}$ and $\tilde{p}_{out}$, randomly sampled in the
regime 
of detectability. The value of the
 normalized mutual information (NMI) between retrieved and ground-truth
 community structure is represented by the color of the various
 points. The green line in the plot identifies combinations
of $p_{in}$ and $p_{out}$ compatible with the observed average degree
$\langle k \rangle$ of the graph.  The blue line is  $y=x$, and denotes the region where
 community structure  is present.  The orange line is the detectability threshold
$\frac{N(Q-1)}{Q}(p_{in} - p_{out}) = \sqrt{\frac{N(Q-1)}{Q}(p_{in} -
  p_{out})}$. }
	\label{heatmap}
\end{figure}

In Fig.~\ref{heatmap}, we display the outcome of our tests when the community
detection algorithm is applied relying on prior information
given by $\tilde{p}_{in}$ and $\tilde{p}_{out}$.
We consider only combinations $(\tilde{p}_{in}, \tilde{p}_{out})$
that lay in the regime of detectability~\cite{decelle2011inference}. 
The figure shows that our algorithm 
reproduces accurately the community structure of the
graph for several combinations $(\tilde{p}_{in}, \tilde{p}_{out})$.
This fact happens as long as  $(\tilde{p}_{in}, \tilde{p}_{out})$ is not too far 
from  the ground truth $(\hat{p}_{in}, \hat{p}_{out})$.
The finding tells us that knowing the exact value is not a necessary
requirement  for the correct detection of the modules; we need only a good guess of
the values of the parameters. In particular, the analysis suggests a
simple criterion for the choice  of the parameter values $p_{in}$ and
$p_{out}$ that can be used in the algorithm.  We can use any combination
that
satisfy the equations
\begin{eqnarray} 
	\label{Ps}
	\begin{split}
		p_{in} + p_{out} &=  \frac{2 \langle k \rangle}{N}\\
		p_{in} - p_{out} &> \frac{2\sqrt{\langle k \rangle}}{N}
	\end{split} \; .
\end{eqnarray}
where $\langle k \rangle$ is the average degree observed in the
network. The first equation imposes that the parameters $p_{in}$ and
$p_{out}$ are compatible with the average degree of the observed network.
The inequality appearing in the bottom of Eq.~(\ref{Ps}) is 
instead restricting our possibilities only in the regime 
of detectability~\cite{decelle2011asymptotic}. As any point in the
segment determined by Eqs.~(\ref{Ps}) is equivalent in terms of 
performance, the values of the 
parameters $p_{in}$ and $p_{out}$ used by our algorithm 
are obtained with
\begin{eqnarray} 
	\label{Ps-algorithms}
	\begin{split}
		p_{in} &= \alpha\frac{\langle k \rangle + \sqrt{\langle k \rangle}}{N} \\
		p_{out} &= \max{ \left\{0, \frac{2 \langle k
                    \rangle}{N} - p_{in} \right\} } \; ,
	\end{split}
\end{eqnarray}

where $\alpha > 0$ is a tunable parameter, whose value is chosen
appropriately such that $p_{in} > p_{out} \geq 0$. In our numerical
results, we set $\alpha = 1.2$. However, we 
verified that the performance of the algorithm doesn't change 
if we choose small $\alpha$ values at random.

\section{Results}

\subsection{Artificial graphs}

First, we perform tests of the 
original and reduced 
versions of the algorithm on synthetic graphs
with pre-imposed community structure. \change{These are compared with
  100 realizations from both the well-established methods 
Louvain~\cite{blondel2008fast} and
  Infomap~\cite{rosvall2008maps}.} \change{In our numerical tests, we used the
implementations of the two algorithms provided by the Python library
{\it igraph}~\cite{igraph}.} \change{In particular, we use the best
partition found by Louvain, the community structure obtained looking at
the lowest level of the multiresolution method~\cite{lancichinetti2009community}.}
 We consider two different 
variants of the stochastic block model: the
Girvan-Newman (GN) benchmark graph~\cite{girvan2002community} and the
Lancichinetti-Fortunato-Radicchi (LFR) benchmark graph~\cite{lancichinetti2008benchmark}.
We measure the performance of the algorithms using NMI as a function
of the community strength of the model, determined by the value of the
mixing parameter 
$\mu = \frac{k_{out}}{k_{out} +  k_{in}}$, i.e., the ratio between external
and total degree of the nodes. This parametrization allows
for a direct comparison between our results on those 
reported in Ref.~\cite{lancichinetti2009community}.

\begin{figure}[!htb]
	\centering
	\includegraphics[width=.5\textwidth]{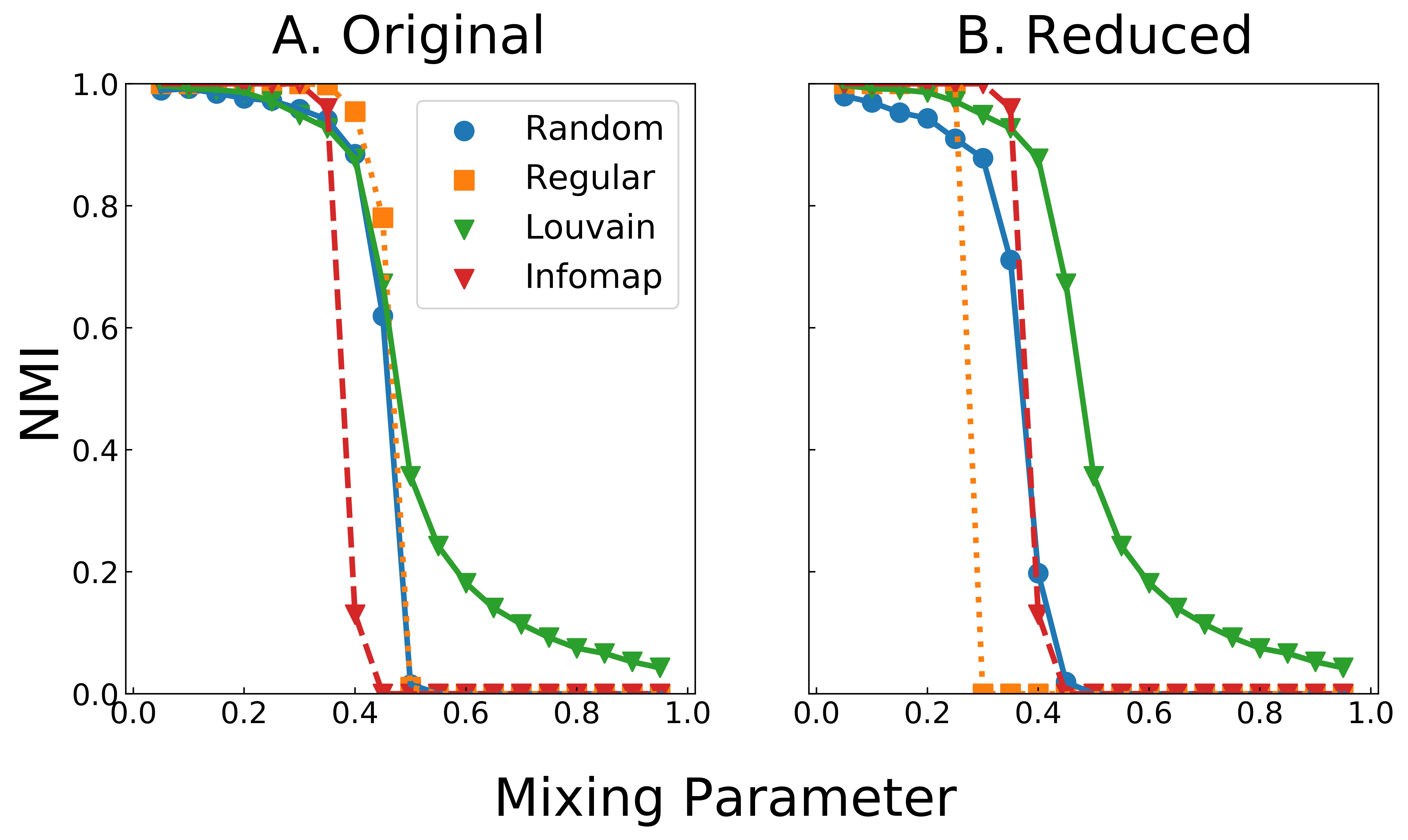}
	\caption{ Performance of the community detection algorithm on
          the Girvan-Newman (GN) benchmark graph. We plot measure values of
          the normalized mutual information (NMI) as a function of the
          mixing parameter $\mu$ of the model.
          A) Results of the
          original version of the
		algorithm with both starting conditions; B)
                Performance of the reduced version of the algorithm. Both of these are compared with Louvain and Infomap. }
	\label{fig:GN}
\end{figure}

In Fig.~\ref{fig:GN}A, we show the performance of the algorithms
on the Girvan-Newman (GN) graph.
The original algorithm is tested on $100$ instances
for each $\mu$ value. We compare results using both starting
conditions. Similarly, Fig.~\ref{fig:GN}B shows the results of the reduced algorithm on $100$ instances of the GN graph. In the original implementation, at around $\mu = 0.5$, the performances of both algorithm reduce to 0. \change{Both tend to outperform Infomap for large values of  $\mu$ but perform worse than Louvain. In the reduced version of the algorithm,
the performance of the regular implementation reduces to 0 when $\mu
\geq 0.3$. The random implementation is similar to Infomap and both
start to drop around $\mu = 0.4$. As before, both perform worse than
Louvain.} The values of $\mu$ where we see a drop in performance
are tantamount with the level of fuzziness where most
of the algorithms start to systematically fail on the GN
benchmark~\cite{lancichinetti2009community}. In most of the cases, 
either perfect communities or one 
large community was predicted. An interesting finding is that 
the reduced version of the algorithm is able to
perform just as well as the original version with the regular
conditions and just 
slightly worse with the random conditions for low values of $\mu$.

\begin{figure}[!htb]
	\centering
	\includegraphics[width=.5\textwidth]{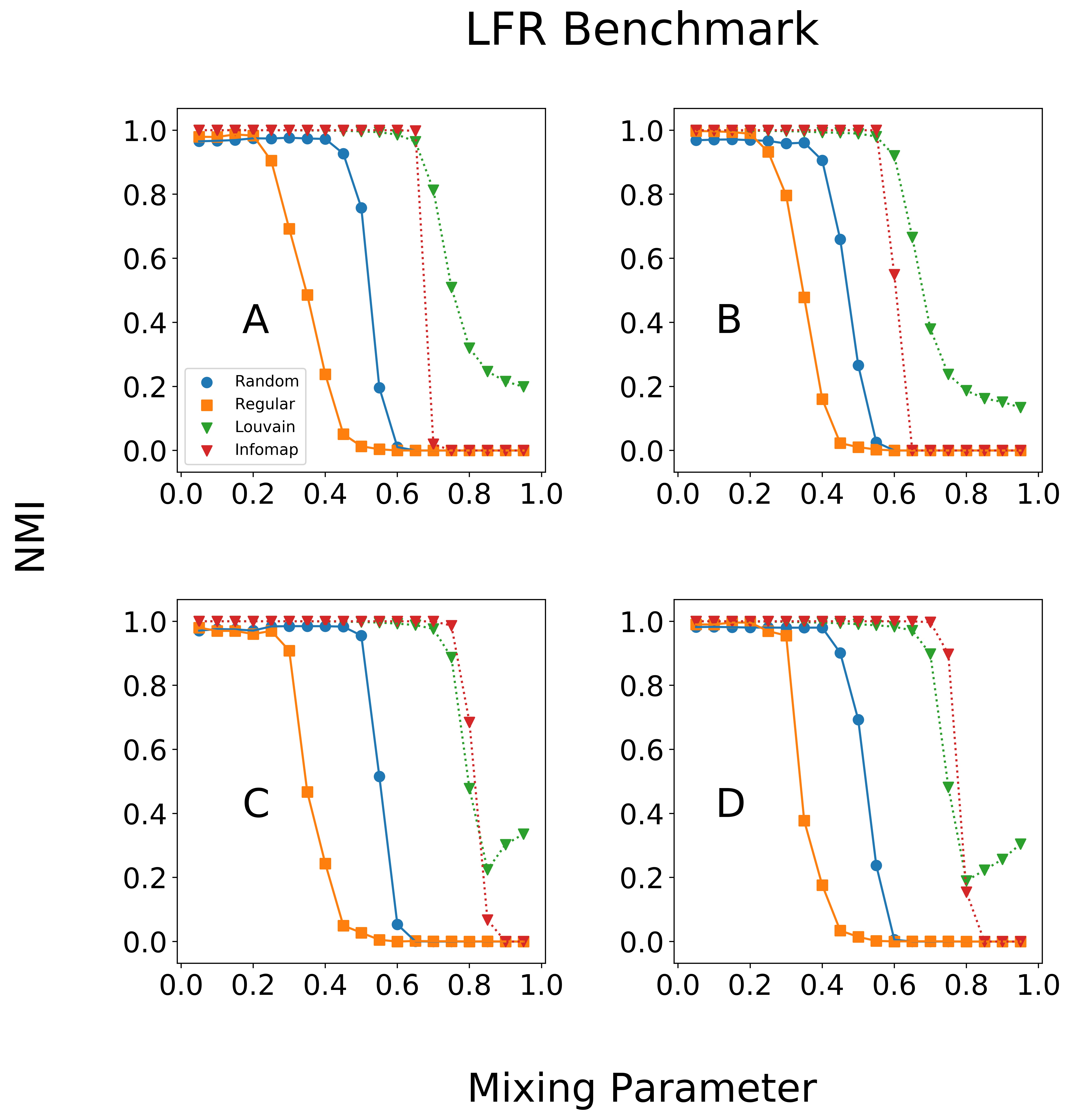}
	\caption{Performance of the community detection algorithm on
          the Lancichinetti-Fortunato-Radicchi (LFR) benchmark
          graph. We display results only for the reduced version of the
          algorithm but with both initial conditions. \change{As a term of
          comparison, we display results obtained by Louvain and
          Infomap in the same set of benchmark graphs.}
          In the various panels, performance is measured in terms of
          normalized mutual information (NMI).
          This quantity is evaluated as a function of the
          mixing parameter $\mu$ of the model. We consider the
          following experimental settings:
	 A) Small communities with $N = 1000$ nodes; B) Big communities
		with $N = 1000$ nodes; C) $N = 5000$ nodes with Small communities;
		D) $N = 5000$ nodes with Big communities. 
	}
	\label{fig:LFR}
\end{figure}

Tests on the LFR graphs are reported in Fig.~\ref{fig:LFR}. Similar to
Ref.~\cite{lancichinetti2009community}, our tests were performed on
networks with size either $N=1000$ or $N=5000$, generated under
condition S, i.e., small
communities with size in the range $[10,50]$ nodes per community, or
under condition $B$, i.e., large
communities with size in the range  $[20, 100]$. In the
  generation of graph instances, community sizes are
chosen at random according to power-law functions with exponent
$-1$ defined over the aforementioned ranges. Node degrees are 
random variates extracted from a power-law degree distribution
with exponent $-2$, such that the average degree of the
nodes is $20$ and maximum degree equals $50$.
We tested the performance of our algorithms over 
$100$ instances of the model for each $\mu$ value. 
Given the high complexity of the original version of the algorithm, we
could test in a systematic fashion only the 
performance of the reduced algorithm. The algorithm was 
started from both initial
conditions.  The results of Fig.~\ref{fig:LFR} provide evidence that
the algorithm is able to achieve good performance, although the
ability to recover the right community structure of the model
decreases to zero for a level of noise slightly smaller than those
of other algorithms~\cite{lancichinetti2009community}. 

\subsection{Real networks}

Recently, community detection algorithms have been focusing on
incorporating edge and 
node metadata into community formation~\cite{newman_structure_2016}. 
An interesting point in this context is understanding how much the
community structure of a network is actually representative 
for exogenous classifications of nodes obtainable from
metadata~\cite{hric_community_2014}.

We run both versions of the algorithms 100 times on 5 well-known
datasets with metadata. For each dataset, we applied three filters; splitting communities into connected components, removing duplicates, and removing singletons~\cite{hric_community_2014}. The Zachary Karate Club network is a social
network of 34 nodes and 78 edges of self reported
friends~\cite{zachary1977information}. A disagreement between the two
leaders led to the splitting of the club into two groups. The US
College football network is a network of college football teams in
which edges represent a scheduled game in the Fall of
2000~\cite{girvan2002community}. The communities are the 12
conferences each of the teams belong to. 
\change{The US Political Book network represents all books
  co-purchased on Amazon.com around the 2004 election in which edges are Amazon recommendations indicating co-purchases from other users while the groups represent the political leanings of the book (Liberal, Neutral, or Conservative) found by human ratings~\cite{krebs2008network}. The US Political Blog dataset is a network of hyperlinks between blogs with the groups being Conservative or Liberal~\cite{adamic2005political}. Finally, the Facebook social networks are undirected friendship networks from 97 different colleges across the US~\cite{traud2012social}. We specifically use network 82 with dorms, gender, high school, and major as the communities. Due to the size, we only ran 5 iterations on the Facebook network.}

\begin{table*}[!htb]

\begin{tabular}{lrrr|rrr|rrr|rrr|rrr|rr|}
\cline{5-18}
                                        & \multicolumn{1}{l}{}                & \multicolumn{1}{l}{}                & \multicolumn{1}{l|}{}                     & \multicolumn{3}{c|}{\textbf{Original Random}}                                                             & \multicolumn{3}{c|}{\textbf{Original Regular}}                                                            & \multicolumn{3}{c|}{\textbf{Reduced Random}}                                                              & \multicolumn{3}{l|}{\textbf{Reduced Regular}}                                                             & \multicolumn{1}{c|}{\textbf{Louvain}} & \multicolumn{1}{c|}{\textbf{Infomap}} \\ \hline
\multicolumn{1}{|l|}{\textbf{Graphs}}   & \multicolumn{1}{r|}{\textbf{$N$}} & \multicolumn{1}{r|}{\textbf{$M$}} & \multicolumn{1}{r|}{\textbf{$C$}} & \multicolumn{1}{r}{\textbf{5\%}} & \multicolumn{1}{r}{\textbf{50\%}} & \multicolumn{1}{r|}{\textbf{95\%}} & \multicolumn{1}{r}{\textbf{5\%}} & \multicolumn{1}{r}{\textbf{50\%}} & \multicolumn{1}{r|}{\textbf{95\%}} & \multicolumn{1}{r}{\textbf{5\%}} & \multicolumn{1}{r}{\textbf{50\%}} & \multicolumn{1}{r|}{\textbf{95\%}} & \multicolumn{1}{r}{\textbf{5\%}} & \multicolumn{1}{r}{\textbf{50\%}} & \multicolumn{1}{r|}{\textbf{95\%}} & \multicolumn{1}{r|}{\textbf{50\%}}    & \textbf{50\%}                         \\ \hline
\multicolumn{1}{|l|}{\textbf{karate}}   & 34                                  & 78                                  & 2                                         & 0.23                             & 0.43                              & 0.49                               & 0.38                             & 0.47                              & 0.51                               & 0.21                             & 0.68                              & 0.84                               & 0.00                               & 0.00                                 & 0.84                               & 0.52                                  & 0.58                                  \\
\multicolumn{1}{|l|}{\textbf{football}} & 115                                 & 615                                 & 13                                        & 0.72                             & 0.78                              & 0.81                               & 0.86                             & 0.89                              & 0.90                               & 0.86                             & 0.94                              & 0.98                               & 0.34                             & 0.83                              & 0.98                               & 0.21                                  & 0.49                                  \\
\multicolumn{1}{|l|}{\textbf{polbooks}} & 105                                 & 441                                 & 3                                         & 0.42                             & 0.45                              & 0.48                               & 0.43                             & 0.45                              & 0.47                               & 0.27                             & 0.62                              & 0.66                               & 0.60                             & 0.62                              & 0.70                               & 0.04                                  & 0.11                                  \\
\multicolumn{1}{|l|}{\textbf{polblogs}} & 1222                                & 16714                               & 2                                         &                -                  &                   -                &                     -               &                   -               &                     -              &                                    & 0.53                             & 0.59                              & 0.67                               & 0.04                             & 0.78                              & 0.79                               & 0.01                                  & 0.02                                  \\ \hline
\multicolumn{1}{|l|}{\textbf{fb-dorm}} & 10001                               & 362892                              & 112                                       &             -                     &               -                    &                      -              &                   -               &               -                    &              -                      & 0.01                             & 0.11                              & 0.12                               & 0.01                             & 0.06                              & 0.13                               & 0.21                                  & 0.13                                  \\
\multicolumn{1}{|l|}{\textbf{fb-gender}} &                   10001                 &                362892                     & 7                                         &                 -                 &                   -                &                 -                   &                  -                &                    -               &                    -                & 0.00                             & 0.01                              & 0.03                               & 0.01                             & 0.01                              & 0.02                               & 0.00                                  & 0.01                                  \\
\multicolumn{1}{|l|}{\textbf{fb-high school}} &            10001                         &                 362892                    & 691                                       &               -                   &                -                   &               -                     &                   -               &                 -                  &                   -                 & 0.13                             & 0.16                              & 0.17                               & 0.02                             & 0.08                              & 0.16                               & 0.10                                  & 0.29                                  \\
\multicolumn{1}{|l|}{\textbf{fb-major}} &                 10001                    &             362892                        & 180                                       &                 -                 &                  -                 &                       -             &                -                  &                  -                 &                  -                  & 0.01                             & 0.08                              & 0.09                               & 0.01                             & 0.06                              & 0.07                               & 0.03                                  & 0.12                                  \\ \hline
\end{tabular}

\caption{
\change{
NMI of the metadata communities and the 
communities detected by several algorithms applied to five network
datasets: Zachary Karate Club~\cite{zachary1977information}, 2000 US College
football~\cite{girvan2002community}, the 2004 US Political Books from
Amazon~\cite{krebs2008network}, the US Political Blogs~\cite{adamic2005political}, and a
small portion of the Facebook social
network~\cite{traud2012social}. For Facebook we consider different
metadata to define communities. The first four columns report
respectively name of
the network (and eventual metadata used to define communities), number
of nodes $N$ in the network, number of edges $M$, and number of
communities $C$ according to the metadata classification.
All other columns refer to results obtained using community detection algorithms.
First, we considered our
proposed algorithms (original and reduced) and the two different
starting conditions (random and regular). Given the stochasticity of
the outcome, we report median values and the $90\%$ confidence
interval for NMI values. Due to its high computational complexity, we couldn't use the original version of the
our proposed algorithm to analyze large networks. We performed the same analysis using Louvain and
Infomap. In this case, the outcome of the community detection algorithm is
deterministic, so we report a single NMI value.}
}
\label{table:metadata}

\end{table*}

Table~\ref{table:metadata} shows the performance of
algorithms, under both initial conditions, on the various datasets.
Performance is still measured in terms of NMI between
the community structure  recovered by the algorithms and the one
given by the metadata. Best matches between topological 
communities and metadata were
observed for the US College Football network, similar to Hric {\it et
  al.}~\cite{hric_community_2014}. The result is expected as college
football teams 
play more against teams within their conference rather than 
teams outside their conference. 
\change{Interestingly, the communities found by our algorithm seem to
provide significantly higher NMI values than those obtained via
Louvain and Infomap on the US Political Book and US Political Blog networks.}

\section{Conclusion}

In this paper, we exploited the interpretation of the problem of
defining and 
identifying communities in networks as a classical communication 
task over a noisy channel, and made use of a widely-used decoding
technique to generate a novel algorithm for community detection.
Although the primitive version of the algorithm was introduced in
Ref.~\cite{radicchi2017decoding}, we extended the idea
in three respects. First, we generalized the algorithm, originally
designed for the detection of two communities only, to the detection
of an arbitrary number of communities. The generalization consists of
iterating the binary version of the algorithm till
convergence. Second, we accounted for the sparsity of graphs which
community detection methods are usually applied to, and reduced the
complexity of the algorithm from quadratic to linear. The
simplification allowed us to generate a method able to deal with 
potentially large networks without renouncing too much to the basic principles
of the original version of the algorithm. Third, we systematically tested
the performance of the new algorithm on both 
synthetic networks and real-world graphs. These
tests provided 
results that are consistent with what already observed in the
literature for other
well-established algorithms for community detection.
\change{In particular, the algorithm outperformed top community
  detection algorithms in tests based on the standard 
SBM, i.e., involving the detection of equally sized
  communities in graphs with homogenous degree distributions.}
\change{On the basis of the performance results obtained here,
we believe that our algorithm may represent
an effective and efficient alternative to other methods that rely
on the SBM ansatz to
infer network community structure. }

\section*{Availability of data and material}
The code and the datasets used in this analysis are available in a Github repository at https://github.com/kbathina/Error-decoding-community-detection. 

\

\begin{acknowledgements}
FR acknowledges support from the National Science Foundation
        (CMMI-1552487) and from the US Army Research Office (W911NF-16-1-0104). 
\end{acknowledgements}

\bibliographystyle{bmc-mathphys}

\begin{thebibliography}{49}
\ifx \bisbn   \undefined \def \bisbn  #1{ISBN #1}\fi
\ifx \binits  \undefined \def \binits#1{#1}\fi
\ifx \bauthor  \undefined \def \bauthor#1{#1}\fi
\ifx \batitle  \undefined \def \batitle#1{#1}\fi
\ifx \bjtitle  \undefined \def \bjtitle#1{#1}\fi
\ifx \bvolume  \undefined \def \bvolume#1{\textbf{#1}}\fi
\ifx \byear  \undefined \def \byear#1{#1}\fi
\ifx \bissue  \undefined \def \bissue#1{#1}\fi
\ifx \bfpage  \undefined \def \bfpage#1{#1}\fi
\ifx \blpage  \undefined \def \blpage #1{#1}\fi
\ifx \burl  \undefined \def \burl#1{\textsf{#1}}\fi
\ifx \doiurl  \undefined \def \doiurl#1{\textsf{#1}}\fi
\ifx \betal  \undefined \def \betal{\textit{et al.}}\fi
\ifx \binstitute  \undefined \def \binstitute#1{#1}\fi
\ifx \binstitutionaled  \undefined \def \binstitutionaled#1{#1}\fi
\ifx \bctitle  \undefined \def \bctitle#1{#1}\fi
\ifx \beditor  \undefined \def \beditor#1{#1}\fi
\ifx \bpublisher  \undefined \def \bpublisher#1{#1}\fi
\ifx \bbtitle  \undefined \def \bbtitle#1{#1}\fi
\ifx \bedition  \undefined \def \bedition#1{#1}\fi
\ifx \bseriesno  \undefined \def \bseriesno#1{#1}\fi
\ifx \blocation  \undefined \def \blocation#1{#1}\fi
\ifx \bsertitle  \undefined \def \bsertitle#1{#1}\fi
\ifx \bsnm \undefined \def \bsnm#1{#1}\fi
\ifx \bsuffix \undefined \def \bsuffix#1{#1}\fi
\ifx \bparticle \undefined \def \bparticle#1{#1}\fi
\ifx \barticle \undefined \def \barticle#1{#1}\fi
\ifx \bconfdate \undefined \def \bconfdate #1{#1}\fi
\ifx \botherref \undefined \def \botherref #1{#1}\fi
\ifx \url \undefined \def \url#1{\textsf{#1}}\fi
\ifx \bchapter \undefined \def \bchapter#1{#1}\fi
\ifx \bbook \undefined \def \bbook#1{#1}\fi
\ifx \bcomment \undefined \def \bcomment#1{#1}\fi
\ifx \oauthor \undefined \def \oauthor#1{#1}\fi
\ifx \citeauthoryear \undefined \def \citeauthoryear#1{#1}\fi
\ifx \endbibitem  \undefined \def \endbibitem {}\fi
\ifx \bconflocation  \undefined \def \bconflocation#1{#1}\fi
\ifx \arxivurl  \undefined \def \arxivurl#1{\textsf{#1}}\fi
\csname PreBibitemsHook\endcsname

\bibitem{girvan2002community}
\begin{barticle}
\bauthor{\bsnm{Girvan}, \binits{M.}},
\bauthor{\bsnm{Newman}, \binits{M.E.}}:
\batitle{Community structure in social and biological networks}.
\bjtitle{Proceedings of the national academy of sciences}
\bvolume{99}(\bissue{12}),
\bfpage{7821}--\blpage{7826}
(\byear{2002})
\end{barticle}
\endbibitem

\bibitem{fortunato2010community}
\begin{barticle}
\bauthor{\bsnm{Fortunato}, \binits{S.}}:
\batitle{Community detection in graphs}.
\bjtitle{Physics reports}
\bvolume{486}(\bissue{3-5}),
\bfpage{75}--\blpage{174}
(\byear{2010})
\end{barticle}
\endbibitem

\bibitem{friedman2001elements}
\begin{bbook}
\bauthor{\bsnm{Friedman}, \binits{J.}},
\bauthor{\bsnm{Hastie}, \binits{T.}},
\bauthor{\bsnm{Tibshirani}, \binits{R.}}:
\bbtitle{The Elements of Statistical Learning}
vol. \bseriesno{1}.
\bpublisher{Springer}, \blocation{???}
(\byear{2001})
\end{bbook}
\endbibitem

\bibitem{newman2004finding}
\begin{barticle}
\bauthor{\bsnm{Newman}, \binits{M.E.}},
\bauthor{\bsnm{Girvan}, \binits{M.}}:
\batitle{Finding and evaluating community structure in networks}.
\bjtitle{Physical review E}
\bvolume{69}(\bissue{2}),
\bfpage{026113}
(\byear{2004})
\end{barticle}
\endbibitem

\bibitem{newman2004fast}
\begin{barticle}
\bauthor{\bsnm{Newman}, \binits{M.E.}}:
\batitle{Fast algorithm for detecting community structure in networks}.
\bjtitle{Physical review E}
\bvolume{69}(\bissue{6}),
\bfpage{066133}
(\byear{2004})
\end{barticle}
\endbibitem

\bibitem{clauset2004finding}
\begin{barticle}
\bauthor{\bsnm{Clauset}, \binits{A.}},
\bauthor{\bsnm{Newman}, \binits{M.E.}},
\bauthor{, \binits{C.}}:
\batitle{Finding community structure in very large networks}.
\bjtitle{Physical review E}
\bvolume{70}(\bissue{6}),
\bfpage{066111}
(\byear{2004})
\end{barticle}
\endbibitem

\bibitem{guimera2007module}
\begin{barticle}
\bauthor{\bsnm{Guimera}, \binits{R.}},
\bauthor{\bsnm{Sales-Pardo}, \binits{M.}},
\bauthor{\bsnm{Amaral}, \binits{L.A.N.}}:
\batitle{Module identification in bipartite and directed networks}.
\bjtitle{Physical Review E}
\bvolume{76}(\bissue{3}),
\bfpage{036102}
(\byear{2007})
\end{barticle}
\endbibitem

\bibitem{duch2005community}
\begin{barticle}
\bauthor{\bsnm{Duch}, \binits{J.}},
\bauthor{\bsnm{Arenas}, \binits{A.}}:
\batitle{Community detection in complex networks using extremal optimization}.
\bjtitle{Physical review E}
\bvolume{72}(\bissue{2}),
\bfpage{027104}
(\byear{2005})
\end{barticle}
\endbibitem

\bibitem{newman2006finding}
\begin{barticle}
\bauthor{\bsnm{Newman}, \binits{M.E.}}:
\batitle{Finding community structure in networks using the eigenvectors of
  matrices}.
\bjtitle{Physical review E}
\bvolume{74}(\bissue{3}),
\bfpage{036104}
(\byear{2006})
\end{barticle}
\endbibitem

\bibitem{newman2006modularity}
\begin{barticle}
\bauthor{\bsnm{Newman}, \binits{M.E.}}:
\batitle{Modularity and community structure in networks}.
\bjtitle{Proceedings of the national academy of sciences}
\bvolume{103}(\bissue{23}),
\bfpage{8577}--\blpage{8582}
(\byear{2006})
\end{barticle}
\endbibitem

\bibitem{zhou2003distance}
\begin{barticle}
\bauthor{\bsnm{Zhou}, \binits{H.}}:
\batitle{Distance, dissimilarity index, and network community structure}.
\bjtitle{Physical review e}
\bvolume{67}(\bissue{6}),
\bfpage{061901}
(\byear{2003})
\end{barticle}
\endbibitem

\bibitem{rosvall2008maps}
\begin{barticle}
\bauthor{\bsnm{Rosvall}, \binits{M.}},
\bauthor{\bsnm{Bergstrom}, \binits{C.T.}}:
\batitle{Maps of random walks on complex networks reveal community structure}.
\bjtitle{Proceedings of the National Academy of Sciences}
\bvolume{105}(\bissue{4}),
\bfpage{1118}--\blpage{1123}
(\byear{2008})
\end{barticle}
\endbibitem

\bibitem{newman2007mixture}
\begin{barticle}
\bauthor{\bsnm{Newman}, \binits{M.E.}},
\bauthor{\bsnm{Leicht}, \binits{E.A.}}:
\batitle{Mixture models and exploratory analysis in networks}.
\bjtitle{Proceedings of the National Academy of Sciences}
\bvolume{104}(\bissue{23}),
\bfpage{9564}--\blpage{9569}
(\byear{2007})
\end{barticle}
\endbibitem

\bibitem{hastings2006community}
\begin{barticle}
\bauthor{\bsnm{Hastings}, \binits{M.B.}}:
\batitle{Community detection as an inference problem}.
\bjtitle{Physical Review E}
\bvolume{74}(\bissue{3}),
\bfpage{035102}
(\byear{2006})
\end{barticle}
\endbibitem

\bibitem{decelle2011inference}
\begin{barticle}
\bauthor{\bsnm{Decelle}, \binits{A.}},
\bauthor{\bsnm{Krzakala}, \binits{F.}},
\bauthor{\bsnm{Moore}, \binits{C.}},
\bauthor{\bsnm{Zdeborov{\'a}}, \binits{L.}}:
\batitle{Inference and phase transitions in the detection of modules in sparse
  networks}.
\bjtitle{Physical Review Letters}
\bvolume{107}(\bissue{6}),
\bfpage{065701}
(\byear{2011})
\end{barticle}
\endbibitem

\bibitem{karrer2011stochastic}
\begin{barticle}
\bauthor{\bsnm{Karrer}, \binits{B.}},
\bauthor{\bsnm{Newman}, \binits{M.E.}}:
\batitle{Stochastic blockmodels and community structure in networks}.
\bjtitle{Physical review E}
\bvolume{83}(\bissue{1}),
\bfpage{016107}
(\byear{2011})
\end{barticle}
\endbibitem

\bibitem{peixoto2014hierarchical}
\begin{barticle}
\bauthor{\bsnm{Peixoto}, \binits{T.P.}}:
\batitle{Hierarchical block structures and high-resolution model selection in
  large networks}.
\bjtitle{Physical Review X}
\bvolume{4}(\bissue{1}),
\bfpage{011047}
(\byear{2014})
\end{barticle}
\endbibitem

\bibitem{peixoto2013parsimonious}
\begin{barticle}
\bauthor{\bsnm{Peixoto}, \binits{T.P.}}:
\batitle{Parsimonious module inference in large networks}.
\bjtitle{Physical review letters}
\bvolume{110}(\bissue{14}),
\bfpage{148701}
(\byear{2013})
\end{barticle}
\endbibitem

\bibitem{peixoto2017bayesian}
\begin{botherref}
\oauthor{\bsnm{Peixoto}, \binits{T.P.}}:
Bayesian stochastic blockmodeling.
arXiv preprint arXiv:1705.10225
(2017)
\end{botherref}
\endbibitem

\bibitem{peel2017ground}
\begin{barticle}
\bauthor{\bsnm{Peel}, \binits{L.}},
\bauthor{\bsnm{Larremore}, \binits{D.B.}},
\bauthor{\bsnm{Clauset}, \binits{A.}}:
\batitle{The ground truth about metadata and community detection in networks}.
\bjtitle{Science advances}
\bvolume{3}(\bissue{5}),
\bfpage{1602548}
(\byear{2017})
\end{barticle}
\endbibitem

\bibitem{nadakuditi2012graph}
\begin{barticle}
\bauthor{\bsnm{Nadakuditi}, \binits{R.R.}},
\bauthor{\bsnm{Newman}, \binits{M.E.}}:
\batitle{Graph spectra and the detectability of community structure in
  networks}.
\bjtitle{Physical review letters}
\bvolume{108}(\bissue{18}),
\bfpage{188701}
(\byear{2012})
\end{barticle}
\endbibitem

\bibitem{krzakala2013spectral}
\begin{barticle}
\bauthor{\bsnm{Krzakala}, \binits{F.}},
\bauthor{\bsnm{Moore}, \binits{C.}},
\bauthor{\bsnm{Mossel}, \binits{E.}},
\bauthor{\bsnm{Neeman}, \binits{J.}},
\bauthor{\bsnm{Sly}, \binits{A.}},
\bauthor{\bsnm{Zdeborov{\'a}}, \binits{L.}},
\bauthor{\bsnm{Zhang}, \binits{P.}}:
\batitle{Spectral redemption in clustering sparse networks}.
\bjtitle{Proceedings of the National Academy of Sciences}
\bvolume{110}(\bissue{52}),
\bfpage{20935}--\blpage{20940}
(\byear{2013})
\end{barticle}
\endbibitem

\bibitem{radicchi2013detectability}
\begin{barticle}
\bauthor{\bsnm{Radicchi}, \binits{F.}}:
\batitle{Detectability of communities in heterogeneous networks}.
\bjtitle{Physical Review E}
\bvolume{88}(\bissue{1}),
\bfpage{010801}
(\byear{2013})
\end{barticle}
\endbibitem

\bibitem{radicchi2014paradox}
\begin{barticle}
\bauthor{\bsnm{Radicchi}, \binits{F.}}:
\batitle{A paradox in community detection}.
\bjtitle{EPL (Europhysics Letters)}
\bvolume{106}(\bissue{3}),
\bfpage{38001}
(\byear{2014})
\end{barticle}
\endbibitem

\bibitem{abbe2015community}
\begin{bchapter}
\bauthor{\bsnm{Abbe}, \binits{E.}},
\bauthor{\bsnm{Sandon}, \binits{C.}}:
\bctitle{Community detection in general stochastic block models: Fundamental
  limits and efficient algorithms for recovery}.
In: \bbtitle{Foundations of Computer Science (FOCS), 2015 IEEE 56th Annual
  Symposium On},
pp. \bfpage{670}--\blpage{688}
(\byear{2015}).
\bcomment{IEEE}
\end{bchapter}
\endbibitem

\bibitem{abbe2017community}
\begin{botherref}
\oauthor{\bsnm{Abbe}, \binits{E.}}:
Community detection and stochastic block models: recent developments.
arXiv preprint arXiv:1703.10146
(2017)
\end{botherref}
\endbibitem

\bibitem{abbe2016exact}
\begin{barticle}
\bauthor{\bsnm{Abbe}, \binits{E.}},
\bauthor{\bsnm{Bandeira}, \binits{A.S.}},
\bauthor{\bsnm{Hall}, \binits{G.}}:
\batitle{Exact recovery in the stochastic block model}.
\bjtitle{IEEE Transactions on Information Theory}
\bvolume{62}(\bissue{1}),
\bfpage{471}--\blpage{487}
(\byear{2016})
\end{barticle}
\endbibitem

\bibitem{mossel2013proof}
\begin{botherref}
\oauthor{\bsnm{Mossel}, \binits{E.}},
\oauthor{\bsnm{Neeman}, \binits{J.}},
\oauthor{\bsnm{Sly}, \binits{A.}}:
A proof of the block model threshold conjecture.
Combinatorica,
1--44
(2013)
\end{botherref}
\endbibitem

\bibitem{shannon2001mathematical}
\begin{barticle}
\bauthor{\bsnm{Shannon}, \binits{C.E.}}:
\batitle{A mathematical theory of communication}.
\bjtitle{ACM SIGMOBILE mobile computing and communications review}
\bvolume{5}(\bissue{1}),
\bfpage{3}--\blpage{55}
(\byear{2001})
\end{barticle}
\endbibitem

\bibitem{radicchi2017decoding}
\begin{botherref}
\oauthor{\bsnm{Radicchi}, \binits{F.}}:
Decoding communities in networks.
arXiv preprint arXiv:1711.05310
(2017)
\end{botherref}
\endbibitem

\bibitem{gallager1962low}
\begin{barticle}
\bauthor{\bsnm{Gallager}, \binits{R.}}:
\batitle{Low-density parity-check codes}.
\bjtitle{IRE Transactions on information theory}
\bvolume{8}(\bissue{1}),
\bfpage{21}--\blpage{28}
(\byear{1962})
\end{barticle}
\endbibitem

\bibitem{mackay1996near}
\begin{barticle}
\bauthor{\bsnm{MacKay}, \binits{D.J.}},
\bauthor{\bsnm{Neal}, \binits{R.M.}}:
\batitle{Near shannon limit performance of low density parity check codes}.
\bjtitle{Electronics letters}
\bvolume{32}(\bissue{18}),
\bfpage{1645}
(\byear{1996})
\end{barticle}
\endbibitem

\bibitem{newman2013community}
\begin{barticle}
\bauthor{\bsnm{Newman}, \binits{M.E.}}:
\batitle{Community detection and graph partitioning}.
\bjtitle{EPL (Europhysics Letters)}
\bvolume{103}(\bissue{2}),
\bfpage{28003}
(\byear{2013})
\end{barticle}
\endbibitem

\bibitem{kernighan1970efficient}
\begin{barticle}
\bauthor{\bsnm{Kernighan}, \binits{B.W.}},
\bauthor{\bsnm{Lin}, \binits{S.}}:
\batitle{An efficient heuristic procedure for partitioning graphs}.
\bjtitle{The Bell system technical journal}
\bvolume{49}(\bissue{2}),
\bfpage{291}--\blpage{307}
(\byear{1970})
\end{barticle}
\endbibitem

\bibitem{fiduccia1982linear}
\begin{bchapter}
\bauthor{\bsnm{Fiduccia}, \binits{C.M.}},
\bauthor{\bsnm{Mattheyses}, \binits{R.M.}}:
\bctitle{A linear-time heuristic for improving network partitions}.
In: \bbtitle{Proceedings of the 19th Design Automation Conference},
pp. \bfpage{175}--\blpage{181}
(\byear{1982}).
\bcomment{IEEE Press}
\end{bchapter}
\endbibitem

\bibitem{mackay2003information}
\begin{bbook}
\bauthor{\bsnm{MacKay}, \binits{D.J.}},
\bauthor{\bsnm{Mac~Kay}, \binits{D.J.}}:
\bbtitle{Information Theory, Inference and Learning Algorithms}.
\bpublisher{Cambridge university press}, \blocation{???}
(\byear{2003})
\end{bbook}
\endbibitem

\bibitem{strehl2002cluster}
\begin{barticle}
\bauthor{\bsnm{Strehl}, \binits{A.}},
\bauthor{\bsnm{Ghosh}, \binits{J.}}:
\batitle{Cluster ensembles---a knowledge reuse framework for combining multiple
  partitions}.
\bjtitle{Journal of machine learning research}
\bvolume{3}(\bissue{Dec}),
\bfpage{583}--\blpage{617}
(\byear{2002})
\end{barticle}
\endbibitem

\bibitem{danon2005comparing}
\begin{barticle}
\bauthor{\bsnm{Danon}, \binits{L.}},
\bauthor{\bsnm{Diaz-Guilera}, \binits{A.}},
\bauthor{\bsnm{Duch}, \binits{J.}},
\bauthor{\bsnm{Arenas}, \binits{A.}}:
\batitle{Comparing community structure identification}.
\bjtitle{Journal of Statistical Mechanics: Theory and Experiment}
\bvolume{2005}(\bissue{09}),
\bfpage{09008}
(\byear{2005})
\end{barticle}
\endbibitem

\bibitem{decelle2011asymptotic}
\begin{barticle}
\bauthor{\bsnm{Decelle}, \binits{A.}},
\bauthor{\bsnm{Krzakala}, \binits{F.}},
\bauthor{\bsnm{Moore}, \binits{C.}},
\bauthor{\bsnm{Zdeborov{\'a}}, \binits{L.}}:
\batitle{Asymptotic analysis of the stochastic block model for modular networks
  and its algorithmic applications}.
\bjtitle{Physical Review E}
\bvolume{84}(\bissue{6}),
\bfpage{066106}
(\byear{2011})
\end{barticle}
\endbibitem

\bibitem{blondel2008fast}
\begin{barticle}
\bauthor{\bsnm{Blondel}, \binits{V.D.}},
\bauthor{\bsnm{Guillaume}, \binits{J.-L.}},
\bauthor{\bsnm{Lambiotte}, \binits{R.}},
\bauthor{\bsnm{Lefebvre}, \binits{E.}}:
\batitle{Fast unfolding of communities in large networks}.
\bjtitle{Journal of statistical mechanics: theory and experiment}
\bvolume{2008}(\bissue{10}),
\bfpage{10008}
(\byear{2008})
\end{barticle}
\endbibitem

\bibitem{igraph}
\begin{botherref}
python-igraph.
\url{ http://igraph.org/python}.
Accessed: 2019-01-10
\end{botherref}
\endbibitem

\bibitem{lancichinetti2009community}
\begin{barticle}
\bauthor{\bsnm{Lancichinetti}, \binits{A.}},
\bauthor{\bsnm{Fortunato}, \binits{S.}}:
\batitle{Community detection algorithms: a comparative analysis}.
\bjtitle{Physical review E}
\bvolume{80}(\bissue{5}),
\bfpage{056117}
(\byear{2009})
\end{barticle}
\endbibitem

\bibitem{lancichinetti2008benchmark}
\begin{barticle}
\bauthor{\bsnm{Lancichinetti}, \binits{A.}},
\bauthor{\bsnm{Fortunato}, \binits{S.}},
\bauthor{\bsnm{Radicchi}, \binits{F.}}:
\batitle{Benchmark graphs for testing community detection algorithms}.
\bjtitle{Physical review E}
\bvolume{78}(\bissue{4}),
\bfpage{046110}
(\byear{2008})
\end{barticle}
\endbibitem

\bibitem{newman_structure_2016}
\begin{barticle}
\bauthor{\bsnm{Newman}, \binits{M.E.J.}},
\bauthor{\bsnm{Clauset}, \binits{A.}}:
\batitle{Structure and inference in annotated networks}.
\bjtitle{Nature Communications}
\bvolume{7},
\bfpage{11863}
(\byear{2016}).
doi:\doiurl{10.1038/ncomms11863}.
Accessed 2017-11-08
\end{barticle}
\endbibitem

\bibitem{hric_community_2014}
\begin{barticle}
\bauthor{\bsnm{Hric}, \binits{D.}},
\bauthor{\bsnm{Darst}, \binits{R.K.}},
\bauthor{\bsnm{Fortunato}, \binits{S.}}:
\batitle{Community detection in networks: {Structural} communities versus
  ground truth}.
\bjtitle{Physical Review E}
\bvolume{90}(\bissue{6}),
\bfpage{062805}
(\byear{2014}).
doi:\doiurl{10.1103/PhysRevE.90.062805}
\end{barticle}
\endbibitem

\bibitem{zachary1977information}
\begin{barticle}
\bauthor{\bsnm{Zachary}, \binits{W.W.}}:
\batitle{An information flow model for conflict and fission in small groups}.
\bjtitle{Journal of anthropological research}
\bvolume{33}(\bissue{4}),
\bfpage{452}--\blpage{473}
(\byear{1977})
\end{barticle}
\endbibitem

\bibitem{krebs2008network}
\begin{botherref}
\oauthor{\bsnm{Krebs}, \binits{V.}}:
A network of books about recent us politics sold by the online bookseller
  amazon. com.
Unpublished http://www. orgnet. com
(2008)
\end{botherref}
\endbibitem

\bibitem{adamic2005political}
\begin{bchapter}
\bauthor{\bsnm{Adamic}, \binits{L.A.}},
\bauthor{\bsnm{Glance}, \binits{N.}}:
\bctitle{The political blogosphere and the 2004 us election: divided they
  blog}.
In: \bbtitle{Proceedings of the 3rd International Workshop on Link Discovery},
pp. \bfpage{36}--\blpage{43}
(\byear{2005}).
\bcomment{ACM}
\end{bchapter}
\endbibitem

\bibitem{traud2012social}
\begin{barticle}
\bauthor{\bsnm{Traud}, \binits{A.L.}},
\bauthor{\bsnm{Mucha}, \binits{P.J.}},
\bauthor{\bsnm{Porter}, \binits{M.A.}}:
\batitle{Social structure of facebook networks}.
\bjtitle{Physica A: Statistical Mechanics and its Applications}
\bvolume{391}(\bissue{16}),
\bfpage{4165}--\blpage{4180}
(\byear{2012})
\end{barticle}
\endbibitem

\end{thebibliography}


\newcommand{\BMCxmlcomment}[1]{}

\BMCxmlcomment{

<refgrp>

<bibl id="B1">
  <title><p>Community structure in social and biological networks</p></title>
  <aug>
    <au><snm>Girvan</snm><fnm>M</fnm></au>
    <au><snm>Newman</snm><fnm>ME</fnm></au>
  </aug>
  <source>Proceedings of the national academy of sciences</source>
  <publisher>National Acad Sciences</publisher>
  <pubdate>2002</pubdate>
  <volume>99</volume>
  <issue>12</issue>
  <fpage>7821</fpage>
  <lpage>-7826</lpage>
</bibl>

<bibl id="B2">
  <title><p>Community detection in graphs</p></title>
  <aug>
    <au><snm>Fortunato</snm><fnm>S</fnm></au>
  </aug>
  <source>Physics reports</source>
  <publisher>Elsevier</publisher>
  <pubdate>2010</pubdate>
  <volume>486</volume>
  <issue>3-5</issue>
  <fpage>75</fpage>
  <lpage>-174</lpage>
</bibl>

<bibl id="B3">
  <title><p>The elements of statistical learning</p></title>
  <aug>
    <au><snm>Friedman</snm><fnm>J</fnm></au>
    <au><snm>Hastie</snm><fnm>T</fnm></au>
    <au><snm>Tibshirani</snm><fnm>R</fnm></au>
  </aug>
  <publisher>Springer series in statistics New York, NY, USA:</publisher>
  <pubdate>2001</pubdate>
  <volume>1</volume>
  <issue>10</issue>
</bibl>

<bibl id="B4">
  <title><p>Finding and evaluating community structure in networks</p></title>
  <aug>
    <au><snm>Newman</snm><fnm>ME</fnm></au>
    <au><snm>Girvan</snm><fnm>M</fnm></au>
  </aug>
  <source>Physical review E</source>
  <publisher>APS</publisher>
  <pubdate>2004</pubdate>
  <volume>69</volume>
  <issue>2</issue>
  <fpage>026113</fpage>
</bibl>

<bibl id="B5">
  <title><p>Fast algorithm for detecting community structure in
  networks</p></title>
  <aug>
    <au><snm>Newman</snm><fnm>ME</fnm></au>
  </aug>
  <source>Physical review E</source>
  <publisher>APS</publisher>
  <pubdate>2004</pubdate>
  <volume>69</volume>
  <issue>6</issue>
  <fpage>066133</fpage>
</bibl>

<bibl id="B6">
  <title><p>Finding community structure in very large networks</p></title>
  <aug>
    <au><snm>Clauset</snm><fnm>A</fnm></au>
    <au><snm>Newman</snm><fnm>ME</fnm></au>
    <au><fnm>C</fnm></au>
  </aug>
  <source>Physical review E</source>
  <publisher>APS</publisher>
  <pubdate>2004</pubdate>
  <volume>70</volume>
  <issue>6</issue>
  <fpage>066111</fpage>
</bibl>

<bibl id="B7">
  <title><p>Module identification in bipartite and directed
  networks</p></title>
  <aug>
    <au><snm>Guimera</snm><fnm>R</fnm></au>
    <au><snm>Sales Pardo</snm><fnm>M</fnm></au>
    <au><snm>Amaral</snm><fnm>LAN</fnm></au>
  </aug>
  <source>Physical Review E</source>
  <publisher>APS</publisher>
  <pubdate>2007</pubdate>
  <volume>76</volume>
  <issue>3</issue>
  <fpage>036102</fpage>
</bibl>

<bibl id="B8">
  <title><p>Community detection in complex networks using extremal
  optimization</p></title>
  <aug>
    <au><snm>Duch</snm><fnm>J</fnm></au>
    <au><snm>Arenas</snm><fnm>A</fnm></au>
  </aug>
  <source>Physical review E</source>
  <publisher>APS</publisher>
  <pubdate>2005</pubdate>
  <volume>72</volume>
  <issue>2</issue>
  <fpage>027104</fpage>
</bibl>

<bibl id="B9">
  <title><p>Finding community structure in networks using the eigenvectors of
  matrices</p></title>
  <aug>
    <au><snm>Newman</snm><fnm>ME</fnm></au>
  </aug>
  <source>Physical review E</source>
  <publisher>APS</publisher>
  <pubdate>2006</pubdate>
  <volume>74</volume>
  <issue>3</issue>
  <fpage>036104</fpage>
</bibl>

<bibl id="B10">
  <title><p>Modularity and community structure in networks</p></title>
  <aug>
    <au><snm>Newman</snm><fnm>ME</fnm></au>
  </aug>
  <source>Proceedings of the national academy of sciences</source>
  <publisher>National Acad Sciences</publisher>
  <pubdate>2006</pubdate>
  <volume>103</volume>
  <issue>23</issue>
  <fpage>8577</fpage>
  <lpage>-8582</lpage>
</bibl>

<bibl id="B11">
  <title><p>Distance, dissimilarity index, and network community
  structure</p></title>
  <aug>
    <au><snm>Zhou</snm><fnm>H</fnm></au>
  </aug>
  <source>Physical review e</source>
  <publisher>APS</publisher>
  <pubdate>2003</pubdate>
  <volume>67</volume>
  <issue>6</issue>
  <fpage>061901</fpage>
</bibl>

<bibl id="B12">
  <title><p>Maps of random walks on complex networks reveal community
  structure</p></title>
  <aug>
    <au><snm>Rosvall</snm><fnm>M</fnm></au>
    <au><snm>Bergstrom</snm><fnm>CT</fnm></au>
  </aug>
  <source>Proceedings of the National Academy of Sciences</source>
  <publisher>National Acad Sciences</publisher>
  <pubdate>2008</pubdate>
  <volume>105</volume>
  <issue>4</issue>
  <fpage>1118</fpage>
  <lpage>-1123</lpage>
</bibl>

<bibl id="B13">
  <title><p>Mixture models and exploratory analysis in networks</p></title>
  <aug>
    <au><snm>Newman</snm><fnm>ME</fnm></au>
    <au><snm>Leicht</snm><fnm>EA</fnm></au>
  </aug>
  <source>Proceedings of the National Academy of Sciences</source>
  <publisher>National Acad Sciences</publisher>
  <pubdate>2007</pubdate>
  <volume>104</volume>
  <issue>23</issue>
  <fpage>9564</fpage>
  <lpage>-9569</lpage>
</bibl>

<bibl id="B14">
  <title><p>Community detection as an inference problem</p></title>
  <aug>
    <au><snm>Hastings</snm><fnm>MB</fnm></au>
  </aug>
  <source>Physical Review E</source>
  <publisher>APS</publisher>
  <pubdate>2006</pubdate>
  <volume>74</volume>
  <issue>3</issue>
  <fpage>035102</fpage>
</bibl>

<bibl id="B15">
  <title><p>Inference and phase transitions in the detection of modules in
  sparse networks</p></title>
  <aug>
    <au><snm>Decelle</snm><fnm>A</fnm></au>
    <au><snm>Krzakala</snm><fnm>F</fnm></au>
    <au><snm>Moore</snm><fnm>C</fnm></au>
    <au><snm>Zdeborov{\'a}</snm><fnm>L</fnm></au>
  </aug>
  <source>Physical Review Letters</source>
  <publisher>APS</publisher>
  <pubdate>2011</pubdate>
  <volume>107</volume>
  <issue>6</issue>
  <fpage>065701</fpage>
</bibl>

<bibl id="B16">
  <title><p>Stochastic blockmodels and community structure in
  networks</p></title>
  <aug>
    <au><snm>Karrer</snm><fnm>B</fnm></au>
    <au><snm>Newman</snm><fnm>ME</fnm></au>
  </aug>
  <source>Physical review E</source>
  <publisher>APS</publisher>
  <pubdate>2011</pubdate>
  <volume>83</volume>
  <issue>1</issue>
  <fpage>016107</fpage>
</bibl>

<bibl id="B17">
  <title><p>Hierarchical block structures and high-resolution model selection
  in large networks</p></title>
  <aug>
    <au><snm>Peixoto</snm><fnm>TP</fnm></au>
  </aug>
  <source>Physical Review X</source>
  <publisher>APS</publisher>
  <pubdate>2014</pubdate>
  <volume>4</volume>
  <issue>1</issue>
  <fpage>011047</fpage>
</bibl>

<bibl id="B18">
  <title><p>Parsimonious module inference in large networks</p></title>
  <aug>
    <au><snm>Peixoto</snm><fnm>TP</fnm></au>
  </aug>
  <source>Physical review letters</source>
  <publisher>APS</publisher>
  <pubdate>2013</pubdate>
  <volume>110</volume>
  <issue>14</issue>
  <fpage>148701</fpage>
</bibl>

<bibl id="B19">
  <title><p>Bayesian stochastic blockmodeling</p></title>
  <aug>
    <au><snm>Peixoto</snm><fnm>TP</fnm></au>
  </aug>
  <source>arXiv preprint arXiv:1705.10225</source>
  <pubdate>2017</pubdate>
</bibl>

<bibl id="B20">
  <title><p>The ground truth about metadata and community detection in
  networks</p></title>
  <aug>
    <au><snm>Peel</snm><fnm>L</fnm></au>
    <au><snm>Larremore</snm><fnm>DB</fnm></au>
    <au><snm>Clauset</snm><fnm>A</fnm></au>
  </aug>
  <source>Science advances</source>
  <publisher>American Association for the Advancement of Science</publisher>
  <pubdate>2017</pubdate>
  <volume>3</volume>
  <issue>5</issue>
  <fpage>e1602548</fpage>
</bibl>

<bibl id="B21">
  <title><p>Graph spectra and the detectability of community structure in
  networks</p></title>
  <aug>
    <au><snm>Nadakuditi</snm><fnm>RR</fnm></au>
    <au><snm>Newman</snm><fnm>ME</fnm></au>
  </aug>
  <source>Physical review letters</source>
  <publisher>APS</publisher>
  <pubdate>2012</pubdate>
  <volume>108</volume>
  <issue>18</issue>
  <fpage>188701</fpage>
</bibl>

<bibl id="B22">
  <title><p>Spectral redemption in clustering sparse networks</p></title>
  <aug>
    <au><snm>Krzakala</snm><fnm>F</fnm></au>
    <au><snm>Moore</snm><fnm>C</fnm></au>
    <au><snm>Mossel</snm><fnm>E</fnm></au>
    <au><snm>Neeman</snm><fnm>J</fnm></au>
    <au><snm>Sly</snm><fnm>A</fnm></au>
    <au><snm>Zdeborov{\'a}</snm><fnm>L</fnm></au>
    <au><snm>Zhang</snm><fnm>P</fnm></au>
  </aug>
  <source>Proceedings of the National Academy of Sciences</source>
  <publisher>National Acad Sciences</publisher>
  <pubdate>2013</pubdate>
  <volume>110</volume>
  <issue>52</issue>
  <fpage>20935</fpage>
  <lpage>-20940</lpage>
</bibl>

<bibl id="B23">
  <title><p>Detectability of communities in heterogeneous networks</p></title>
  <aug>
    <au><snm>Radicchi</snm><fnm>F</fnm></au>
  </aug>
  <source>Physical Review E</source>
  <publisher>APS</publisher>
  <pubdate>2013</pubdate>
  <volume>88</volume>
  <issue>1</issue>
  <fpage>010801</fpage>
</bibl>

<bibl id="B24">
  <title><p>A paradox in community detection</p></title>
  <aug>
    <au><snm>Radicchi</snm><fnm>F</fnm></au>
  </aug>
  <source>EPL (Europhysics Letters)</source>
  <publisher>IOP Publishing</publisher>
  <pubdate>2014</pubdate>
  <volume>106</volume>
  <issue>3</issue>
  <fpage>38001</fpage>
</bibl>

<bibl id="B25">
  <title><p>Community detection in general stochastic block models: Fundamental
  limits and efficient algorithms for recovery</p></title>
  <aug>
    <au><snm>Abbe</snm><fnm>E</fnm></au>
    <au><snm>Sandon</snm><fnm>C</fnm></au>
  </aug>
  <source>Foundations of Computer Science (FOCS), 2015 IEEE 56th Annual
  Symposium on</source>
  <pubdate>2015</pubdate>
  <fpage>670</fpage>
  <lpage>-688</lpage>
</bibl>

<bibl id="B26">
  <title><p>Community detection and stochastic block models: recent
  developments</p></title>
  <aug>
    <au><snm>Abbe</snm><fnm>E</fnm></au>
  </aug>
  <source>arXiv preprint arXiv:1703.10146</source>
  <pubdate>2017</pubdate>
</bibl>

<bibl id="B27">
  <title><p>Exact recovery in the stochastic block model</p></title>
  <aug>
    <au><snm>Abbe</snm><fnm>E</fnm></au>
    <au><snm>Bandeira</snm><fnm>AS</fnm></au>
    <au><snm>Hall</snm><fnm>G</fnm></au>
  </aug>
  <source>IEEE Transactions on Information Theory</source>
  <publisher>IEEE</publisher>
  <pubdate>2016</pubdate>
  <volume>62</volume>
  <issue>1</issue>
  <fpage>471</fpage>
  <lpage>-487</lpage>
</bibl>

<bibl id="B28">
  <title><p>A proof of the block model threshold conjecture</p></title>
  <aug>
    <au><snm>Mossel</snm><fnm>E</fnm></au>
    <au><snm>Neeman</snm><fnm>J</fnm></au>
    <au><snm>Sly</snm><fnm>A</fnm></au>
  </aug>
  <source>Combinatorica</source>
  <publisher>Springer</publisher>
  <pubdate>2013</pubdate>
  <fpage>1</fpage>
  <lpage>-44</lpage>
</bibl>

<bibl id="B29">
  <title><p>A mathematical theory of communication</p></title>
  <aug>
    <au><snm>Shannon</snm><fnm>CE</fnm></au>
  </aug>
  <source>ACM SIGMOBILE mobile computing and communications review</source>
  <publisher>ACM</publisher>
  <pubdate>2001</pubdate>
  <volume>5</volume>
  <issue>1</issue>
  <fpage>3</fpage>
  <lpage>-55</lpage>
</bibl>

<bibl id="B30">
  <title><p>Decoding communities in networks</p></title>
  <aug>
    <au><snm>Radicchi</snm><fnm>F</fnm></au>
  </aug>
  <source>arXiv preprint arXiv:1711.05310</source>
  <pubdate>2017</pubdate>
</bibl>

<bibl id="B31">
  <title><p>Low-density parity-check codes</p></title>
  <aug>
    <au><snm>Gallager</snm><fnm>R</fnm></au>
  </aug>
  <source>IRE Transactions on information theory</source>
  <publisher>IEEE</publisher>
  <pubdate>1962</pubdate>
  <volume>8</volume>
  <issue>1</issue>
  <fpage>21</fpage>
  <lpage>-28</lpage>
</bibl>

<bibl id="B32">
  <title><p>Near Shannon limit performance of low density parity check
  codes</p></title>
  <aug>
    <au><snm>MacKay</snm><fnm>DJ</fnm></au>
    <au><snm>Neal</snm><fnm>RM</fnm></au>
  </aug>
  <source>Electronics letters</source>
  <publisher>IET</publisher>
  <pubdate>1996</pubdate>
  <volume>32</volume>
  <issue>18</issue>
  <fpage>1645</fpage>
</bibl>

<bibl id="B33">
  <title><p>Community detection and graph partitioning</p></title>
  <aug>
    <au><snm>Newman</snm><fnm>ME</fnm></au>
  </aug>
  <source>EPL (Europhysics Letters)</source>
  <publisher>IOP Publishing</publisher>
  <pubdate>2013</pubdate>
  <volume>103</volume>
  <issue>2</issue>
  <fpage>28003</fpage>
</bibl>

<bibl id="B34">
  <title><p>An efficient heuristic procedure for partitioning
  graphs</p></title>
  <aug>
    <au><snm>Kernighan</snm><fnm>BW</fnm></au>
    <au><snm>Lin</snm><fnm>S</fnm></au>
  </aug>
  <source>The Bell system technical journal</source>
  <publisher>Nokia Bell Labs</publisher>
  <pubdate>1970</pubdate>
  <volume>49</volume>
  <issue>2</issue>
  <fpage>291</fpage>
  <lpage>-307</lpage>
</bibl>

<bibl id="B35">
  <title><p>A linear-time heuristic for improving network
  partitions</p></title>
  <aug>
    <au><snm>Fiduccia</snm><fnm>CM</fnm></au>
    <au><snm>Mattheyses</snm><fnm>RM</fnm></au>
  </aug>
  <source>Proceedings of the 19th design automation conference</source>
  <pubdate>1982</pubdate>
  <fpage>175</fpage>
  <lpage>-181</lpage>
</bibl>

<bibl id="B36">
  <title><p>Information theory, inference and learning algorithms</p></title>
  <aug>
    <au><snm>MacKay</snm><fnm>DJ</fnm></au>
    <au><snm>Mac Kay</snm><fnm>DJ</fnm></au>
  </aug>
  <publisher>Cambridge university press</publisher>
  <pubdate>2003</pubdate>
</bibl>

<bibl id="B37">
  <title><p>Cluster ensembles---a knowledge reuse framework for combining
  multiple partitions</p></title>
  <aug>
    <au><snm>Strehl</snm><fnm>A</fnm></au>
    <au><snm>Ghosh</snm><fnm>J</fnm></au>
  </aug>
  <source>Journal of machine learning research</source>
  <pubdate>2002</pubdate>
  <volume>3</volume>
  <issue>Dec</issue>
  <fpage>583</fpage>
  <lpage>-617</lpage>
</bibl>

<bibl id="B38">
  <title><p>Comparing community structure identification</p></title>
  <aug>
    <au><snm>Danon</snm><fnm>L</fnm></au>
    <au><snm>Diaz Guilera</snm><fnm>A</fnm></au>
    <au><snm>Duch</snm><fnm>J</fnm></au>
    <au><snm>Arenas</snm><fnm>A</fnm></au>
  </aug>
  <source>Journal of Statistical Mechanics: Theory and Experiment</source>
  <publisher>IOP Publishing</publisher>
  <pubdate>2005</pubdate>
  <volume>2005</volume>
  <issue>09</issue>
  <fpage>P09008</fpage>
</bibl>

<bibl id="B39">
  <title><p>Asymptotic analysis of the stochastic block model for modular
  networks and its algorithmic applications</p></title>
  <aug>
    <au><snm>Decelle</snm><fnm>A</fnm></au>
    <au><snm>Krzakala</snm><fnm>F</fnm></au>
    <au><snm>Moore</snm><fnm>C</fnm></au>
    <au><snm>Zdeborov{\'a}</snm><fnm>L</fnm></au>
  </aug>
  <source>Physical Review E</source>
  <publisher>APS</publisher>
  <pubdate>2011</pubdate>
  <volume>84</volume>
  <issue>6</issue>
  <fpage>066106</fpage>
</bibl>

<bibl id="B40">
  <title><p>Fast unfolding of communities in large networks</p></title>
  <aug>
    <au><snm>Blondel</snm><fnm>VD</fnm></au>
    <au><snm>Guillaume</snm><fnm>JL</fnm></au>
    <au><snm>Lambiotte</snm><fnm>R</fnm></au>
    <au><snm>Lefebvre</snm><fnm>E</fnm></au>
  </aug>
  <source>Journal of statistical mechanics: theory and experiment</source>
  <publisher>IOP Publishing</publisher>
  <pubdate>2008</pubdate>
  <volume>2008</volume>
  <issue>10</issue>
  <fpage>P10008</fpage>
</bibl>

<bibl id="B41">
  <title><p>python-igraph</p></title>
  <source>\url{ http://igraph.org/python}</source>
  <note>Accessed: 2019-01-10</note>
</bibl>

<bibl id="B42">
  <title><p>Community detection algorithms: a comparative analysis</p></title>
  <aug>
    <au><snm>Lancichinetti</snm><fnm>A</fnm></au>
    <au><snm>Fortunato</snm><fnm>S</fnm></au>
  </aug>
  <source>Physical review E</source>
  <publisher>APS</publisher>
  <pubdate>2009</pubdate>
  <volume>80</volume>
  <issue>5</issue>
  <fpage>056117</fpage>
</bibl>

<bibl id="B43">
  <title><p>Benchmark graphs for testing community detection
  algorithms</p></title>
  <aug>
    <au><snm>Lancichinetti</snm><fnm>A</fnm></au>
    <au><snm>Fortunato</snm><fnm>S</fnm></au>
    <au><snm>Radicchi</snm><fnm>F</fnm></au>
  </aug>
  <source>Physical review E</source>
  <publisher>APS</publisher>
  <pubdate>2008</pubdate>
  <volume>78</volume>
  <issue>4</issue>
  <fpage>046110</fpage>
</bibl>

<bibl id="B44">
  <title><p>Structure and inference in annotated networks</p></title>
  <aug>
    <au><snm>Newman</snm><fnm>M. E. J.</fnm></au>
    <au><snm>Clauset</snm><fnm>A</fnm></au>
  </aug>
  <source>Nature Communications</source>
  <pubdate>2016</pubdate>
  <volume>7</volume>
  <fpage>ncomms11863</fpage>
  <url>https://www.nature.com/articles/ncomms11863</url>
</bibl>

<bibl id="B45">
  <title><p>Community detection in networks: {Structural} communities versus
  ground truth</p></title>
  <aug>
    <au><snm>Hric</snm><fnm>D</fnm></au>
    <au><snm>Darst</snm><fnm>RK</fnm></au>
    <au><snm>Fortunato</snm><fnm>S</fnm></au>
  </aug>
  <source>Physical Review E</source>
  <pubdate>2014</pubdate>
  <volume>90</volume>
  <issue>6</issue>
  <fpage>062805</fpage>
  <url>https://link.aps.org/doi/10.1103/PhysRevE.90.062805</url>
</bibl>

<bibl id="B46">
  <title><p>An information flow model for conflict and fission in small
  groups</p></title>
  <aug>
    <au><snm>Zachary</snm><fnm>WW</fnm></au>
  </aug>
  <source>Journal of anthropological research</source>
  <publisher>University of New Mexico</publisher>
  <pubdate>1977</pubdate>
  <volume>33</volume>
  <issue>4</issue>
  <fpage>452</fpage>
  <lpage>-473</lpage>
</bibl>

<bibl id="B47">
  <title><p>A network of books about recent US politics sold by the online
  bookseller amazon. com</p></title>
  <aug>
    <au><snm>Krebs</snm><fnm>V</fnm></au>
  </aug>
  <source>Unpublished http://www. orgnet. com</source>
  <pubdate>2008</pubdate>
</bibl>

<bibl id="B48">
  <title><p>The political blogosphere and the 2004 US election: divided they
  blog</p></title>
  <aug>
    <au><snm>Adamic</snm><fnm>LA</fnm></au>
    <au><snm>Glance</snm><fnm>N</fnm></au>
  </aug>
  <source>Proceedings of the 3rd international workshop on Link
  discovery</source>
  <pubdate>2005</pubdate>
  <fpage>36</fpage>
  <lpage>-43</lpage>
</bibl>

<bibl id="B49">
  <title><p>Social structure of Facebook networks</p></title>
  <aug>
    <au><snm>Traud</snm><fnm>AL</fnm></au>
    <au><snm>Mucha</snm><fnm>PJ</fnm></au>
    <au><snm>Porter</snm><fnm>MA</fnm></au>
  </aug>
  <source>Physica A: Statistical Mechanics and its Applications</source>
  <publisher>Elsevier</publisher>
  <pubdate>2012</pubdate>
  <volume>391</volume>
  <issue>16</issue>
  <fpage>4165</fpage>
  <lpage>-4180</lpage>
</bibl>

</refgrp>
} 


\end{document}